\documentclass[utf8]{frontiersSCNS}  
\usepackage{url,hyperref,lineno,microtype,subcaption}
\usepackage[onehalfspacing]{setspace}

\def\keyFont{\fontsize{8}{11}\helveticabold }
\def\firstAuthorLast{Heeralal Janwa {et~al.}}

\def\Authors{Heeralal Janwa\,$^{1,*}$, Steven E. Massey\,$^{2}$, Julian Velev\,$^{3}$ and Bud Mishra\,$^{4}$}

\def\Address{
$^{1}$ Department of Mathematics, University of Puerto Rico, San Juan, PR 00931 \\
$^{2}$ Department of Biology, University of Puerto Rico, San Juan, PR 00931\\ 
$^{3}$ Department of Physics, University of Puerto Rico, San Juan, PR 00931 \\ 
$^{4}$ Departments of Computer Science, Mathematics and Cell Biology, Courant Institute and NYU School of Medicine, New York University , NY
}

\def\corrAuthor{Bud Mishra} \def\corrEmail{mishra@nyu.edu}

\begin{document}
\onecolumn
\firstpage{1}

\title[Origin of Biomolecular Networks]{Origin of Biomolecular Networks} 

\author[\firstAuthorLast ]{\Authors} 
\address{} 
\correspondance{} 

\extraAuth{}

\maketitle

\begin{abstract}

\section{}
Biomolecular networks have already found great utility in characterizing complex biological systems arising from pair-wise interactions amongst biomolecules. Here, we review how graph theoretical approaches can be applied not only for a better understanding of various proximate (mechanistic) relations, but also, ultimate (evolutionary) structures encoded in such networks. A central question deals with the evolutionary dynamics by which different topologies of biomolecular networks might have evolved, as well as the biological principles that can be hypothesized from a deeper understanding of the induced network dynamics. We emphasize the role of gene duplication in terms of signaling game theory, whereby sender and receiver gene players accrue benefit from gene duplication, leading to a preferential attachment mode of network growth. Information asymmetry between sender and receiver genes is hypothesized as a key driver of the resulting network topology. The study of the resulting dynamics  suggests many mathematical/computational problems, the majority of which are intractable but yield to efficient approximation algorithms, when studied through an algebraic graph theoretic lens.

\tiny
\keyFont{ \section{Keywords:} Biomolecules, Regulation and Communication, Interaction (Binary) Relationship, Network Model, Network Analysis, Spectral analysis} 
\end{abstract}

\section{Genesis of Biomolecular Interactions}\label{sec:Genesis}

\subsection{Introduction and a Road Map}\label{subsec:Introduction}

A range of complex phenotypes of biomolecular systems can be inferred from macromolecular interactions, represented using networks. Such biomolecular networks include gene (regulatory) networks (GRNs) \cite{genereg}, protein-protein interaction  (PPI) networks \cite{ppi}, protein and RNA neutral networks \cite{neutrna} \cite{neutprot}, metabolic networks \cite{metab} and meta-metabolic networks (composite metabolic networks of communities) \cite{meta}. Our major focus here will be on GRNs and PPI networks, but the principles outlined are also applicable to the other types of biomolecular networks.

The paper covers the following topics: (i) A brief introduction to biomolecular networks (a topic also covered by other accompanying articles); (ii) A compendium of known results in (algebraic and combinatorial) graph theory ; (iii) Algorithmic (and algebraic) complexity, arising in the study of evolution of networks; (iv) Current state of the field and open problems. The list of open problems focuses largely on the following: How to devise efficient (algebraic) algorithms that can shed important lights on {\em game theoretic models of the evolution of biological interactions\/}, given that they are driven by information asymmetry (leading to duplications, complementation, pseudogenization, etc.). Some of these important mechanisms have been studied qualitatively elsewhere, albeit not mathematically rigorously.


\subsection{Ohno's Evolution by Duplication}\label{subsec:Ohno}
At the genetic level, the growth of a GRN or PPI network is driven by gene mutation: e.g., duplication, translocation, inversion, deletion, short indels, and point mutations, of which duplication plays an outsized role. Susumu Ohno coined the phrase evolution by duplication (EBD) to emphasize this evolutionary dynamic \cite{ohno}. The classic view of molecular evolution is that gene families may expand and contract over evolutionary time due to gene duplication and deletion \cite{demuth}. Here, we wish to present a more complex view, by exploring how biomolecular networks may grow, contract, or alter their topology over time, from the relative dynamic contributions and interactions of their constituent genes and gene families. This evolution is ultimately driven by the process of gene duplication and deletion, which leads to node and edge addition, or removal, from a biomolecular network, respectively. Since such variations in the network alter the phenotypes, over which selection operates, the evolution of networks and their features ultimately captures the essence of Darwinian evolution.

Recently, we introduced a signaling games perspective of biochemistry and molecular evolution \cite{mishmas}. There, we focused on interactions between biological macromolecules, which may be described using the framework of sender-receiver signaling games, where an expressed macromolecule such as a protein or RNA, constitutes a signal sent on behalf of a sender agent (e.g., gene). The signal comprises the three-dimensional conformation and physicochemical properties of the macromolecule. A receiver agent (e.g., a gene product, another macromolecule) may then bind to the signal macromolecule, which produces an action (such as an enzymatic reaction). The action produces utility for the participating agents, sender and receiver, and thereby -- albeit indirectly -- a change in overall fitness of the genome (in evolutionary game theory, utility and fitness are treated as analogous). When there is common interest, the utility is expected to benefit both sender and receiver and their selection, thus driving Darwinian evolution.

Replicator dynamics allow the signaling game to be couched in evolutionary terms \cite{tj1}. Replicator dynamics arise from the increased replication of players with higher utility (fitness). Thus, if a gene has a strategy that results in increased utility, then it will  increase in frequency in a population. For a sender gene this would entail sending a signal that results in an increase in utility, while for a receiver gene this would entail undertaking an action that likewise results in an increase in utility. As already suggested, these dynamics represent a process analogous to Darwinian (adaptive) evolution or positive selection. 

Biomolecular signaling games are sustained by information asymmetry between sender and receiver and so their interactions can be represented using directed graphs. Information asymmetry arises because the receiver is uninformed regarding the identity of the sender gene: it must rely on the expressed signal macromolecule to determine the identity of the sender gene. But, this strategy may be open to deception. Most biomolecular signaling games in the cell are between sender and receiver genes which have perfect common interest. This is so, because they are {\em cellularized\/}, chromosome replication is synchronized and so the genes replicate in concert. Such games are termed `Lewis' signaling games, and rely on honest signaling from sender to the receiver \cite{l}. A biomolecular signaling game is illustrated in Figure 1, part (1).

\begin{figure}[ht]
\begin{center}
 \vspace{40pt}%
\includegraphics[trim = 1cm 2cm 4cm 4cm, clip = false, width=0.7\textwidth]{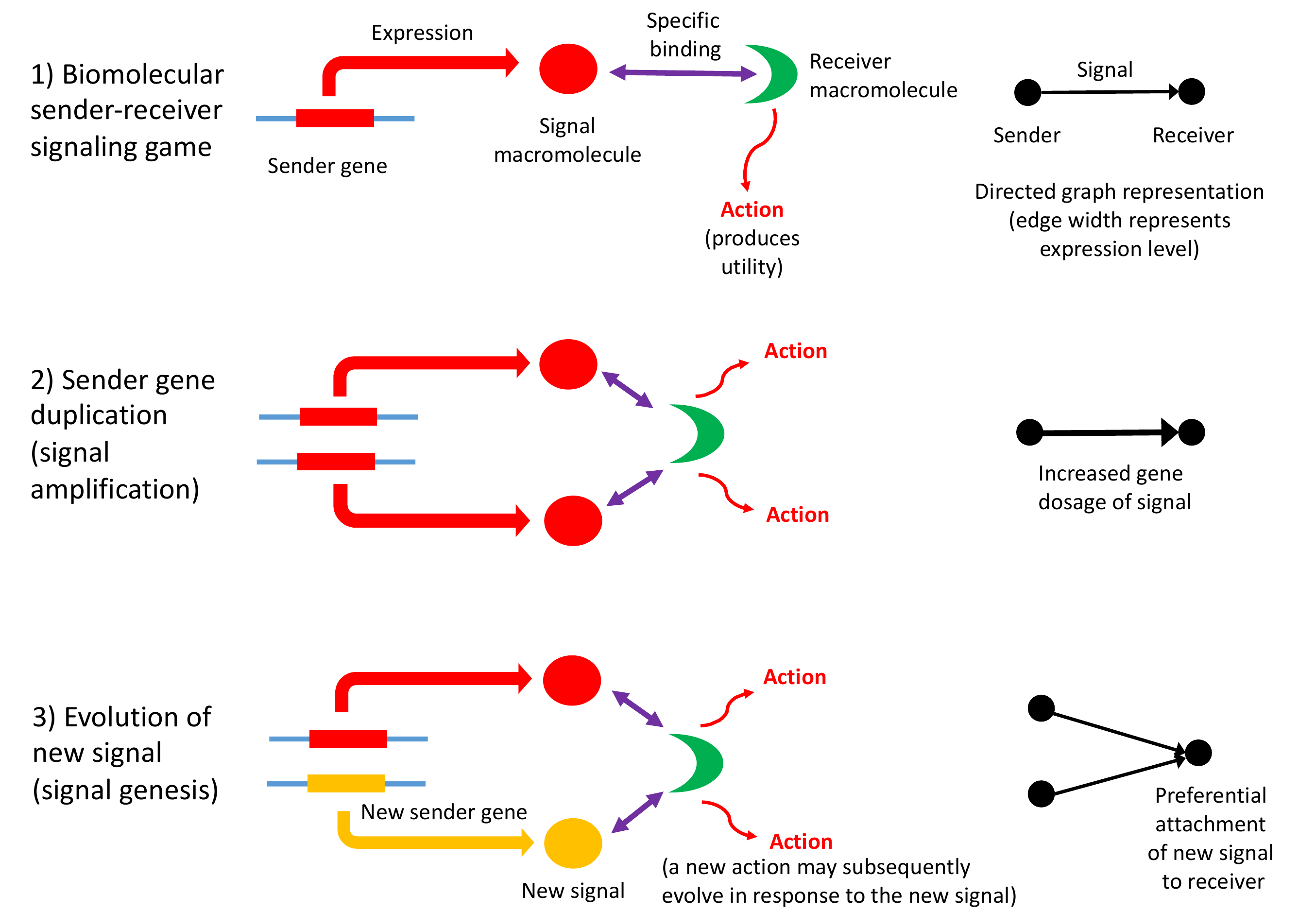}
\end{center}
 \vspace{20pt}%
\caption{{\em The influence of information asymmetry on growth of a PPI network\/}. Interactions between macromolecules are envisaged as a biomolecular signaling game whereby a sender gene expresses a macromolecule, the signal, that then binds specifically to a receiver macromolecule, which then undergoes an action (such as an enzymatic reaction, or conformational change), which produces utility (fitness). The signal consists of the 
three-dimensional conformation and physicochemical properties of the macromolecule (1). The sender gene may undergo duplication, which has a dosage effect on the expressed macromolecule, resulting in signal amplification (2). This mechanism is expected to lower the Shapley value of the gene players in the genome, as the signal is partially redundant and so inefficient. Subsequently, the sender gene duplicate may acquire a new function (evolve a new signal) although the majority would be expected to undergo pseudogenization (3). Both these scenarios represent the re-establishment of a Nash equilibrium. If a new signal macromolecule evolves, it is likely to bind to the same receiver macromolecule initially. This preferential attachment arises because gene duplicates have a tendency to bind to their interaction partner initially, and then subsequently undergo interaction turnover \cite{power}, and is illustrated to the right of the figure. A key problem is how a new action by the receiver arises as the result of the evolution of a new signal; the new action may co-evolve with  the new signal, or may be necessary first before a new signal can evolve. The latter would imply that receiver gene duplication and action genesis facilitates the evolution of new signals and sender genes (an exception would be when there is a conflict of interest; here the sender is more likely to make the first move in evolving a novel deceptive signal, and then the receiver would respond with a better discriminative recognition mechanism). This key, and novel aspect of gene duplication might be deciphered via consideration of the topology of directed graph representations of biomolecular interactions as sender-receiver signaling games. Refinements to the illustrated scheme include situations where the original signal protein binds to a variety of receiver proteins, or where the gene that codes for the receiver protein undergoes duplication (Figure 2). } \label{fig:1}
\end{figure}

However, situations may arise where a sender has a conflict of interest with the receiver. This kind of misalignment can occur when a sender gene is selfish, and would prefer to replicate itself at the expense of the rest of the genome. Such genes are termed `selfish elements,' and come in a variety of forms, all marked by decoupled replication from the rest of the genome \cite{burt}. In a signaling game, when there is a conflict of interest between sender and receiver, then the sender is expected to adopt some degree of deceptive signaling \cite{cs1}. Consistent with this prediction, there are a range of examples of selfish elements that utilize molecular deception \cite{mishmas}.

Gene duplication is a fundamental evolutionary driver of organismal complexity \cite{lwka02}. The first step in the process of duplication of a sender gene may be viewed as one of signal enhancement. Because gene duplication results in gene dosage effects, it also results in amplification of the signal, the expressed gene product. This strategy can be viewed as lowering the overall utility of the genome, given that there is a cost involved in producing excessive signal. It is, thus, expected to lower the Shapley value \cite{shap} of the gene players that cooperate within the genome. This conflict is usually resolved when the duplicated gene becomes pseudogenized, the usual fate of gene duplicates \cite{innan}.

Subsequent to duplication, the gene duplicates will sometimes diverge in function, although the exact mechanism remains to be elucidated \cite{innan}. This process represents signal divergence if the gene is a sender gene, and action divergence if the gene codes for a receiver macromolecule. The genesis of a new sender gene with a new signal may then promote evolution of a novel action by the receiver macromolecule, potentially facilitating duplication of the receiver gene itself. Likewise, the duplication of a receiver gene may facilitate the diversification of macromolecular signals that interact with the two duplicated receiver macromolecules. The process modifies the GRN or PPI network in a non-obvious manner and it deviates considerably from the way evolution of random graphs is usually treated, following Erd{\"o}s and R{\'e}nyi, discussed in more detail in Section 3 \cite{erdos}. 

Signal and action genesis via gene duplication may have features in common with a P\'{o}lya's urn model of signal genesis \cite{skyrm} (P\'{o}lya's urn models are statistical models that involve sampling with replacement influenced by the identity of the sampled element. These models can lead to a `{\em rich get richer\/}' effect, of which 'preferential attachment' is an example, discussed in more detail in Subsection 3.2). In this model, reinforcement of signals (similar to reinforcement learning) may promote the invention of new synonyms. These considerations may provide parallels for how signals originate elsewhere, not dissimilar to how new words in a language can arise from existing words by a process of derivation \cite{word}. Mechanistic commonalities in the process of signal genesis in these diverse systems as exhibited in GRNs remain to be explored. These models hint at a possibly new, but universal form of ``preferential attachment'' that drives the variations in biomolecular networks as well as the selectivity in Darwinian evolution. 

\subsection{Network Topology, Evolution by Duplication,
 and Preferential Attachments}\label{subsec:Network Topology}
Consequently, the topology of gene networks is non-deterministic and yet not memoryless, since it must encode layers of ripples produced earlier via the dynamics of gene duplication (paralogs and orthologs), as amplified during the network's history. Just as physicists infer the theories of origin of universe from the cosmic background radiation, we expect to enrich our understanding of the origin of machinery of life (e.g., codon evolution, evolution of multicellularity, evolution of sex etc.) from a rigorous analysis of the signaling games and their equilibria, which has rippled through the extant biomolecular networks. Taking this analogy further, we observe that the ripples in gravitational waves have been proposed to reflect the existence of parallel universes, whose presence created asymmetries in the initial conditions, giving rise to filamentary structures in the visible universe \cite{hawk} This comparison is inspired by the notion of a `protein big bang' from a single (or handful of) ur-protein(s) in the first complex life forms, evolving by gene duplication into the extant `protein universe,' hinting at the information asymmetries fossilized in the GRN and PPI networks. \cite{bang}.

Likewise, we point out that information asymmetry in macromolecular sender-receiver interactions may point to evolutionary paths that might have been abandoned unexplored; which may suggest new engineering approaches needed by synthetic biology, or in drug discovery, or immuno-therapy. Note that during the process of evolution of signaling, gene duplication and deletion contribute to a certain degree of non-determinism and ``conventionality'' to the Nash equilibria that stabilize and manifest as non-trivial anisotropies in gene network topology.

In summary, the process of gene duplication, tempered by signal and action genesis can be thought of as a driver of preferential attachment in shaping the topology of gene networks, in which information asymmetry between senders and receivers is expected to play an indelible role. Figure 1 illustrates a basic mechanism whereby signal genesis may lead to preferential attachment during the growth of a PPI network. Topological features expected to hint at this process include: $(i)$ the degree distribution, $(ii)$ hierarchicity, $(iii)$ assortativity and many others; they require powerful statistical and algebraic tools -- covered in the later sections, where it is assumed that genome evolution is a complex process involving diverse groups of mutations such as insertions, deletions, conversions, duplications, transpositions, translocations and recombinations, and that it is further affected by selective constraints and effective population size and other factors such as the environment. With recent understanding of large scale cellular networks (regulatory, metabolic, protein-protein interactions) one must now aim at investigation between the evolutionary rates of a gene mutations and its effects on the network topology using mathematical models and analytics: see \cite{wag}. For instance, combining  sequence analysis in a single genome and its close relatives, one can infer the rate and tempo of the evolutionary dynamics acting on the genome, and the resulting effects on the network's algebraic structures. We provide an example of how evolution by duplication leads to a preferential attachment mode of gene network growth in Figure 2, using the duplication of the p53 gene, and its paralogs p63 and p73 -- all  transcription factors regulating pathways involved in related phenotypes of somatic or developmental surveillance and interacting with similar family of genes (e.g., MDM2 or MDMX), as illustration \footnote{A mutation in MDM affects all p53, p63 and p73 allowing utility tradeoffs between fecundity (through increased embryonic lethality) and cancer risks (through reduced somatic surveillance) in a population.}.

\begin{figure}[ht]
\begin{center}
\includegraphics[width=0.7\textwidth]{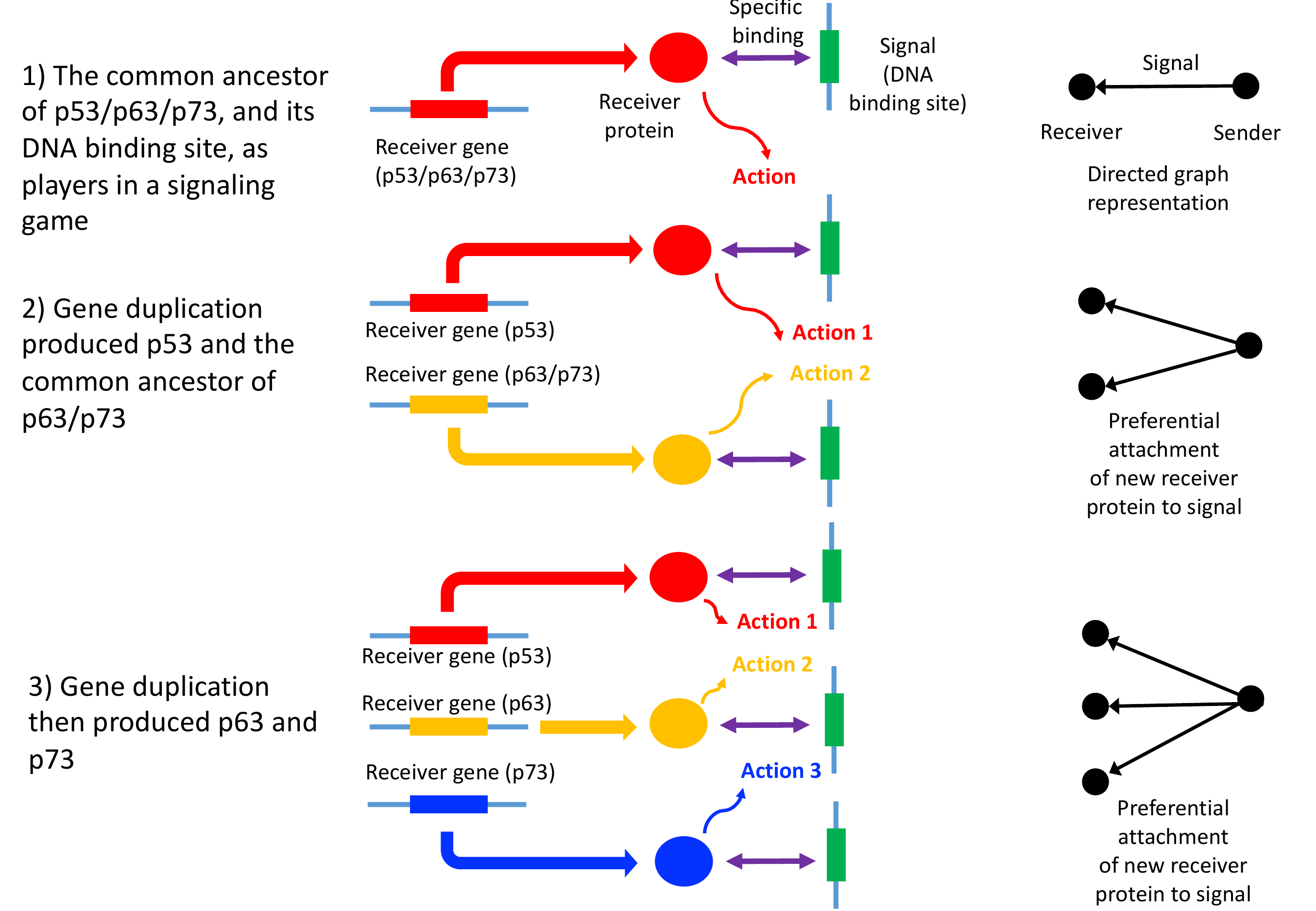}
\end{center}
\caption{{\em Gene duplication of p53, p63 and p73 as a signaling game, and GRN growth.\/} An illustrative example of a signaling games view of network growth is provided by the paralogs p53,  p63 and p73, which code for transcription factors, p53 being of critical importance in many cancers \cite{p53cancer}. Here, p53 and the common ancestor of p63/p73 duplicated (2), followed by the duplication and divergence of p63 and p73 \cite{dup} \cite{p53} (3). The signal is the DNA binding site, while the receivers are the p53, p63 and p73 proteins (here the sender is the protein coding gene downstream of the DNA binding site). The receiver protein undergoes an action upon binding to the DNA binding site (the signal), which consists of the recruitment of additional transcription factors, and contribution to the assembly of the transcription initiation complex \cite{tf}. The gene products of p53, p63 and p73 mostly bind to the same DNA binding sites \cite{binding}, thus each signal (and ultimately sender gene) has acquired two new binding partners, in addition to the original interaction with the gene product of the common ancestor of p53/p63/p73. This is a form of preferential attachment, which should influence network topology as the number of genes increase by duplication, as illustrated to the right of the figure. The signaling games perspective allows us to better understand scenarios where there is a conflict of interest between the genome, and a selfish entity such as a selfish element, a cancer or a virus. When there is a conflict of interest, a deceptive signal is expected to be emitted by the sender \cite{cs1} (the selfish entity). Here, the DNA binding site of the selfish entity will mimic that of canonical DNA binding sites associated with normal cellular function, 'tricking' a transcription factor to bind to it, and altering the transcription of the sender gene (or alternatively abolishing transcription factor binding). Examples include \textit{cis}-regulatory mutations in cancer \cite{cis}} \label{fig:2}
\end{figure}

Note that these abstract models generate refutable hypotheses that need experimental verification and support from mechanistic explanations. However, unfortunately, the biochemical processes involved in the hypothesized preferential attachment dynamics are not fully understood. For example, the duplication processes are often driven by Non-Homologous End Joining (NHEJ), a pathway that repairs double-strand breaks in DNA. To guide repair, NHEJ typically uses short homologous DNA sequences called microhomologies, which are often present in single-stranded overhangs on the ends of double-strand breaks \cite{nhej}. When the overhangs are perfectly compatible, NHEJ usually repairs the break accurately. However, imprecise repair can lead to inappropriate NHEJ resulting in translocations, duplications and rearrangements \cite{imprecise}, which add to variations that are random but not memoryless. Perhaps some of such hypotheses may need to be carefully examined using cancer genome data such as TCGA, and models of tumor progression. This analysis may also explain efficacy of certain therapeutic interventions in cancer as well as their failures via drug and immuno resistance.

\section{Network analysis}\label{sec:Network analysis}

In this section, we discuss fundamentals of graphs, a mathematical formalism used in the study of biomolecular networks, as well as  other related important topics. Consider a set of entities, denoted $V$ and a set of binary relations between the entities $E\ \subseteq V\ \times V$. When $V$ denotes biomolecules and $E$ denotes interactions between them (e.g., regulations, proximity, synteny, etc.), the resulting graph represents a biomolecular network. One important advantage of graphs is that they have intuitive graphical representation. Such networks evolve over time with additions and deletions to the sets $V$ and $E$. In order to create a bridge to algebraic approaches, we extend the standard combinatorial definition by endowing it with additional maps.

Formally, a graph is a pair of sets $G=\left(V,E\right)$ where $V$ are the vertices (nodes, points) and $E\subseteq V\times V$  are the edges (arcs) respectively. When $E$ is a set of unordered pair of vertices the graph is said to be undirected or simple. In a directed graph  $G=\left(V,E,\ o,t\right)$, $E$ consists of an ordered set of vertex pairs, i.e. for each edge $e\in E$, $e\rightarrow\left(o\left(e\right),t\left(e\right)\right)$ where $o\left(e\right)$ is called the origin of the edge $e$ and $t\left(e\right)$ is called the terminus of the edge e [\cite{serre2} and \cite{biggs}]. A graph is weighted if there is a map (weighting function, $w: E \to R_+$) assigning to each edge a positive real-valued weight.

If $G = (V, E, \cdot, \cdot)$ and $G' = (V', E', \cdot, \cdot)$ are two graphs such that $V' \subseteq V$ and $E' \subseteq E \cap (V' \times V')$, then $G' \subseteq G$, $G'$ is a \emph{subgraph} of $G$. If $E' = E \cap (V' \times V')$ ($E'$ contains every edge in $e \in E$ with $o(e), t(e) \in V'$) then $G'$ is an \emph{induced subgraph} of $G$. $G'$ and $G$ are \emph{isomorphic} ($G' \equiv G$) if there is a bijection $f: V' \to V$ with $(u, v) \in E' \iff (f(u), f(v)) \in E$, $\forall u, v \in V'$.

\subsection{Topological Properties}\label{sec:Topological Properties}

Network properties are governed by its topology, such as degree distribution, clustering coefficients, motifs, assortativity, etc. Comprehensive treatments can be found in \cite{thulasiraman2015, Barbasi-book2016}, and for more in-depth treatment regarding biomedical networks in \cite{JLABES-book2017}.

\subsubsection*{Degree Distribution}\label{sec:Degree-distribution}

The degree of a vertex $v$, $\deg(v)$, is the number of edges that connect the vertex with other vertices. In other words, the degree is the number of immediate neighbors of a vertex. In directed graphs in-degree and out-degree of a vertex can be defined as the number of incoming and outgoing edges respectively. Let $n_k$  be the number of vertices of degree $k$ and $|V| = N$, the total number of vertices in the graph and $|E| = M$, the total number of edges in the graph.
 Note that $\sum_k n_k = N$ and $\sum k n_k = \sum_{v \in V} \deg(v) = 2 |E| = 2 M$. The degree distribution is the fraction of vertices of degree $k$, $P(k) = n_k/N$, and two isomorphic networks will have same degree distributions (though not necessarily the converse). Thus, the degree distributions can tell a great deal about the structure of a family of networks. For example, if the degree distribution is singly peaked, following the Poisson (or its Gaussian approximation) distributions, the majority of the nodes can be described by the average degree $\langle k \rangle = \sum_k k P(k) = 2 M/N$. The graph is said to be {\em sparse}, if $\langle k \rangle = o(\log N)$ (or $M = o(N \log N)$). Biomolecular networks are usually sparse, which can be fruitfully exploited in their algorithmic analysis. We can talk of \textit{typical} nodes of the networks as being those that have degree distribution as those within 1 to 2 standard deviations from the average, while, with  probability  decreasing exponentially, it is possible to find nodes with degree much different from the average. While power-law degree distributions follow a completely different pattern: they are {\em fat-tailed}; the majority of the nodes have only few neighbors, while many nodes have relatively large number of neighbors. The highly-connected nodes are known as {\em hubs}.

\subsubsection*{Distance Metrics}\label{subsec:Distance Metrics }

One of the most fundamental metrics is the \emph{distance} on a graph. First we define a \emph{walk} of length $m$ in a graph $G$ from a vertex $u$ to $v$  as a finite alternating sequence of vertices and edges $\langle v_0, e_1, v_1, e_2, \ldots,e_m, v_m \rangle$, such that $o\left(e_i\right)=v_{i-1}$ and $t\left(e_i\right)=v_i$, for $0<i\leq m$,  such that $u=v_0$ and $v=v_m$. Then the number of edges traversed in the shortest walk joining $u$ to $v$ is called the {\it distance} in $G$ between $u$ and $v$ denoted by $d(u,v)$.  If there is a walk from $u$ to itself, then  we say that the set of vertices (respectively edges) form a cycle. The smallest number of $m$ edges in a walk from $u$ to itself is called a cycle of length $m$. The girth $g(G)$, is the shortest cycle in $G$. A walk whose vertices are distinct is called a (simple) \emph{path}.

The concept of a walk allows us to define other properties of the graph. A graph $G=(V,E,\ o,\ e)$ is said to be \emph{connected}, if any two vertices are the extremities of at least one walk. The maximally connected subgraphs are called the {\it connected components} of $G$. A giant component is a connected component containing a significant fraction of the nodes. The maximum value of the distance function in a connected graph is called the {\it diameter} of the graph. Frequently real life networks have small diameter and are said to exhibit  {\it small world phenomenon}. For many biomolecular networks the average distance between two nodes depends logarithmically on the number of vertices in the graph.

Additionally, a {\it complete graph} $G$ is the undirected graph, in which each vertex is a neighbor of all other vertices; $\deg(v) = N -1$, $\forall v \in V$; or equivalently, each distinct pair  of vertices are connected (or are adjacent) by a unique edge. $G$ is then denoted as $K_N$. A {\it clique} in an undirected  graph is a subset of vertices such that its induced subgraph is complete. Additional combinatorial invariants of graphs useful in the analysis of networks can be defined  (see Supplementary material for details).

\subsubsection*{Expanding Constants}\label{subsubsec:Expanding}
Let  ${G}=(V,E, \cdot, \cdot)$ be an undirected graph. Then for all $F\subset V$, the {\it boundary} $\partial F$ is the set of edges connecting $F$ to $V\setminus F$. The {\it expanding constant}, or {\it isoperimetric constant} of ${X}$ is defined as,
$$h({X}) = \min_{\emptyset\not= F\subset V} \frac{|\partial F|} {\min \{ |F|,|V\setminus F| \} }.$$
For molecular network, then, the invariant $h({X})$ measures the quality of the network with respect to the flow of information within it, (e.g., via chemical reactions, or signaling). Larger $h(X)$ implies better expansion, faster mixing, faster partitioning, and many other related properties that may give the network a selective advantage.

Using various combinatorial algorithms devised for the study and analysis of biomolecular networks, one may compute $h(X)$ to determine their complexity. However precise characterization of $h(X)$ itself is an intractable (i.e., NP-complete) problem. Isoperimetric inequalities give bounds on $h(X)$ in terms of a related algebraic invariant, 
 $\gamma(X)$ -- called its \emph{spectral gap}, determination of which has complexity  $O(|V|)^c$, where $c$ is at most $3$; furthermore, $c=1$ for many sparse graphs. We give isoperimetric bounds and  results applicable to biomolecular networks in the Supplement, where we also introduce local Cheeger constant. We also introduce algebraic invariants in Section \ref{sec:Algebraic invariants}.

\subsubsection*{Clustering and Clustering Coefficients}\label{subsubsec:Clustering}


Biological networks  are modular, forming communities and hierarchies, likely to have been sculpted by EBD (Evolution by Duplication). To study these local structures in network science, one may perform {\it community} analysis, which aims to identify a group of nodes that have a higher probability of connecting to each other than to nodes from other communities (see for exmple \cite{PELLEGRINI2019}). Various notions such as $k$-cliques, $k$-clubs and $k$-clans have been developed to detect communities, but they are ultimately closely connected to the problem of finding cliques and consequently, do not generally lend themselves to any reasonable algorithm other than brute-force enumeration. However, even detecting communities approximately may prove valuable for general evolutionary studies, since in these biological networks communities determine how specific biological functions are encoded in cellular networks -- and thus subjected to Darwinian selective pressure, since these players are likely to have formed communities in the first place to carry out specific cellular functions. (see Hartwell, \cite{pmid:10591225}). Figure \ref{fig:Interactome} highlights significant evidence that communities play important role human disease networks (see \cite{JLABES-book2017}).

Usually a simpler approach is commonly employed and deals with the problem of \emph{clustering} in a graph, which seeks to partition the graph into disjoint subgraphs such that nodes in each such subgraph are ``closer'' to the other nodes in the same subgraph, while they are ``farther'' from the nodes of other subgraphs. Hierarchical clustering algorithms have been developed to uncover communities (approximately) in polynomial time and depend upon the {\it similarity matrix} $(x_{ij})$, where the entry $x_{ij}$ equals the distance between node $i$ and node $j$. Among the classical algorithms are included those by Ravasz and by Girvan and Newman \cite{MR1908073}. Other related algorithms include those for random-walk betweenness and network centrality. 
\

The \emph{local clustering coefficient} captures the degree to which the neighbors of a given node link to each other. In general, for undirected graphs, the {\it local  clustering coefficient} $C_i$ of node  $i$ with degree $k_i$ is defined as
$$C_i:= \frac{L_i} {k_i(k_i-1)/2}$$
where the numerator $L_i$ is the actual number of connections between $k_i$ immediate neighbors of $i$, and the denominator is the number of connections if the neighbors formed a complete graph (i.e. a clique). Note that an undirected complete graph $K_{k_i}$ of $k_i$ nodes has $k_i(k_i-1)/2$ edges. Thus, a fully clustered node will have $C_i=1$ and for completely isolated node $C_i=0$. We can define the {\it (average) clustering coefficient} of the whole network with $N$ nodes  as
$$\langle C \rangle = \frac{1}{N} \sum C_i.$$
The clustering coefficients can be used to characterize a network's {\em modularity}, as discussed later (in Section 3) in details.

\subsubsection*{Subgraphs and Motifs}\label{subsubsec:Subgraphs}






Biomolecular networks have been found to contain network {\em motifs}, representing elementary interaction patterns between small subgraphs that occur substantially more often than as predicted by a completely random network of similar size and connectivity. The presence of such motifs is usually explained by an evolutionary process that can quickly create (usually by a variation involving duplication) or eliminate (usually by a selection process that favors pseudogenization and complementation) regulatory interactions in a fast evolutionary time scale -- relative to the rate at which individual genes mutate. It is usually hypothesized that the underlying evolutionary processes are convergent. Thus efficient algorithms to detect such motifs are important in the analysis of biomolecular networks. These algorithms focus on estimating how much more frequently a subgraph isomorphic to a motif graph (with $n$ vertices and $m$ edges) occurs relative to what would be expected by pure chance.


The number $N_{mn}$ of subgraphs with $n$ nodes and $m$ interactions expected of a network of $N$ nodes can be estimated from the two key topological parameters of a complex network -- namely the power-law exponent $\beta$ and the hierarchical exponent $\alpha$ as we discuss in equations (\ref{eqn:powr-law} and \ref{eqn:clustering-power-law})  below. In general the subgraph motifs can be classified in two types: Type I motifs are those where  $(m-n+1)\alpha-(n-\beta)<0$, and type II subgraph motifs are those that satisfy the reverse inequality. One can determine their numbers $N^I$ and $N^{II}$ approximately as a function of $(m-n+1)\alpha-(n-\beta)$ and $n_{max}$, the degree of the most connected node in the network. One can show that $N^I_{nm}>> N^{II}$. One can also show that the relative number of Type II subgraphs is vanishingly small compared to Type I.


\subsection{Algebraic Invariants and Spectrum}\label{sec:Algebraic invariants}

The intuitive pictorial/combinatorial representation of graphs is an extremely useful aid to their  understanding. However, computing the topological properties of graphs combinatorially is computationally challenging especially when the size of the graph becomes large. As noted earlier, indeed, most combinatorial algorithms on biomolecular networks such as on PPI networks and GRNs are computationally complex problems (most of them fall in the NP-complete complexity class) \cite{Karp-2011}. Therefore, in order to carry out any quantitative and computational analysis, graphs are better represented as algebraic objects. This representation allows us to use linear algebra and mathematical analysis techniques. The key to this representation is the adjacency matrix $A(G)$. It is defined as $\{0,1\}^{n\times n}$ matrix in which, $A_{ij}=1$ if the vertices $i$ and $j$ are connected ($\exists e\in E, o(e)=i, t(e)=j)$ and $0$ otherwise. The matrix is symmetric if the graph is undirected. For weighted graphs we can assign weights $w_{ij}$ for existing edges.

Algebraic properties provide us with tools to deduce various properties of the biomolecular networks. In particular, the spectral representation of the graph is of importance for a number of applications such as graph classification, etc. We can think of the adjacency matrix $A$ as operating on the space $V=C^n$ of complex $n$-tuples written as column vectors $x$,$y$ as follows $Ax\rightarrow y$. It can be shown that there are directions left invariant in this space. That is to say, $A_ix_i=\lambda_i x_i$ where $\lambda_i$ are the eigenvalues and corresponding $x_i$ the eigenvectors (spanning invariant directions) of the adjacency matrix for $1\le i\le n$. The spectrum of the graph $G$ is defined as the collection of eigenvalues of the adjacency matrix $\rm{Spec}(G)= \rm{Spec}(A)= {\lambda_1,..,\lambda_n}$. Naturally, if $A$ is a real symmetric matrix, then the eigenvalues of $A$ are real. 


In particular, one algebraic invariant of the graph is the {\it spectral gap} $\gamma(G)$. It can be shown that the spectral gap gives excellent bounds on a combinatorial invariant, the Cheeger constant $h(G)$ (see the Supplementary material).

\section{Network evolution}\label{sec:Network evolution}
Starting with the seminal work of Erd\"os and R\'enyi \cite{erdos}, a number of mathematical frameworks have been developed to model the ``evolution'' of graphs, covering the family of biomolecular networks. These frameworks may prove useful in explaining why most biological networks have certain non-obvious properties: namely, (i) Small-world property; (ii) High clustering coefficients (varying with degree distribution); (iii) Emergence of ``hubs.'' Such network models are ultimately expected to capture various observed properties of biomolecular networks, and the evolutionary trajectories leading up to them.

\subsection{Random Network Models}\label{sec:Random network}

\subsubsection*{Erd\"os and R\'enyi Model}

The Erd\"os and R\'enyi model of random graphs (ER-graphs, denoted $G(n, p)$) is characterized by two parameters, the number of vertices in the network $N$ and the fixed probability of choosing edges $p$ \cite{erdos}. The graph $G$ is generated by choosing $N$ vertices and connecting each pair of vertices with probability $p$. The  model yields a network with approximately $p {\binom{N}{2}} = O(p N^2)$ randomly distributed edges. The probability of choosing a specified graph $G$ with $N$ vertices and $e$ edges is therefore $ \binom{M} {e} p^e(1-p)^{M- e}$, where $M= {\frac{N}{2}}=$ the maximum number of possible edges connecting $N$ vertices.

It can be shown that in such random graph the average vertex degree is $\langle k \rangle = p(N-1) = O(pN)$. The diameter of such graph is $d= \ln N/\ln \langle k \rangle \approx \ln N/ (\ln N - ln (1/p))$ which is small compared to the graph size. Thus, random graphs exhibit ``the small world property.'' The degree distribution for ER graphs is a binomial distribution $P[\deg(u) = k] = \binom{(N-1)}{k} p^k (1-p)^{N-k-1}$, which for large $N$ (relative to $1/p$: where $N = \lambda/p$) converges to the Poisson distribution $P[\deg(u) = k] = e^{-\lambda} \frac{\lambda^k}{k!}$. Then the local clustering coefficient is $C_i= p$ is independent of the degree of the node and the average clustering coefficient $C=p/N$ scales with the network size. Therefore, the standard ER random model seems not to capture either the properties of degree distribution or the clustering coefficient of biomolecular networks.

Typically, an ER random graph model is used as a ``null model'' for the evolutionary process. However, while deviations from randomness are frequently used as evidence for the direct action of natural selection, often non-randomness may reflect neutrally generated (non-adaptive) emergent phenomena \cite{nemerge}. We emphasize here that many topological features of biomolecular networks are unlikely to be directly selected for, but instead are a side-product of network growth, and decay, captured by the dynamics of edge and node addition and removal.

\subsubsection*{Small World Model}\label{sec:Small world model}

The biomolecular networks have features that are not captured by the Erd\"os and R\'enyi random graph model. As we have seen, random graphs have low clustering  coefficient and they do not account for formation of hubs. To rectify some of these shortcomings, the {\it small world model} or popularly known as the {\it six degree of separation model} was introduced as the next level of complexity for probabilistic model with features that are  closer to the real world networks  \cite{Watts1998CollectiveDO, MR1716136}. The evolution and dynamics of such networks have been discussed in detail  \cite{MR2041642}, in particular in the diseases propagation literature~\cite{MR2125834}.

In this model the graph $G$ of $N$ nodes is constructed as a ring lattice, in which, (i) first, \emph{wire}: that is, connect every node to $K/2$ neighbors on each side and (ii) second, \emph{rewire}: that is, for every edge connecting a particular node, with probability $p$ reconnect it to a randomly selected node.

The average number of such edges is $pNK/2$. The first step of the algorithm produces local clustering, while the second dramatically reduces the distance in the network. Unlike the random graph, the clustering coefficient of this network $C= 3(K-2)/4(K-1)$ is independent of the system size. Thus, the small world network model displays the small world property and the clustering of real networks, however, it does not capture the emergence of hubby nodes (e.g., P53 in biomolecular networks).

\subsection{Scale-free Network Models}\label{sec:Scale-free network models}

Most biomolecular networks are  hypothesized to have a degree distribution, described as \emph{scale-free}. In a scale free network the number of nodes $n_k$ of degree $k$ is proportional to a power of the degree, namely, the degree distribution of the nodes follows a \emph{power-law} 
\begin{equation}
{n_k} = k^{-\beta}, \label{eqn:powr-law}
\end{equation}
where $\beta>1$ is a coefficient characteristic of the network \cite{MR2091634}. Unlike in random networks, where the degree of all nodes is centered around a single value -- with the probability of finding nodes with much larger (or smaller) degree decaying exponentially, in scale-free networks there are nodes of large degree with relatively higher probability (\emph{fat tail\/}). In other words, since the power low distribution decreases much more slowly than exponential, for large $k$ (heavy or fat tails), scale-free networks support nodes with extremely high number of connections called ``hubs.'' Power law distribution has been observed in many large networks, such as the Internet, the phone-call maps, the collaboration networks, etc. \cite{Kepes-book-2007, MR2548299, Barbasi-book2016}. A caveat to these reports is that inappropriate statistical techniques are often been used to infer power law distributions, and alternative heavy tailed distributions may fit the data better \cite{clau}.  However, the power law is a useful approximation that allows mechanisms of network growth to be explored, such as Preferential Attachment, discussed next, while the examination of alternative heavy tailed distributions is set as an Open Problem.

\subsubsection*{Preferential Attachment}\label{subsubsec:Preferential}

The original model of \emph{preferential attachment} was proposed by Barabási–Albert \cite{MR2091634}. The scheme consists of a local {\it growth rule} that leads to a global consequence, namely a power law distribution. The network grows through the addition of new nodes linking to nodes already present in the system. There is higher probability to preferentially link to a node with a large number of connections. Thus, this rule gives more preferences to those vertices that have larger degrees. For this reason it is often referred to as the ``rich-get-richer'' or ``Matthew'' effect. 

With an initial  graph $G_0$ and a fixed probability parameter $p$, the preferential attachment random graph model $G(p,G_0)$ can be described as follows: at each step the graph $G_t$ is formed by modifying the earlier graph $G_{t-1}$ in two steps -- with probability $p$ take a \emph{vertex-step}; otherwise, take an \emph{edge-step}:
\begin{itemize}
\item[(i)] {\it Vertex step:} Add a new vertex $v$ and an edge $\{u,v\}$ from $v$ to $u$ by randomly and independently choosing $u$ proportional its degree;
\item[(ii)] {\it Edge step:} Add a new edge $\{r,s\}$ by independently choosing vertices $r$ and $s$ with probability proportional to their degrees.
\end{itemize}
That is, at each step, we add a vertex with probability $p$, while for sure, we add an additional edge. If we denote by $n_t$ and $e_t$ the number of vertices and edges respectively at step $t$, then $e_t=t+1$ and $n_t=1+\sum_{i=1}^t z_i$, where $z_i$'s are Bernoulli random variables with probability of success $=p$. Hence the expected value of nodes is $\langle n_t \rangle=1+pt$.

It can be shown that exponentially (as $t$ asymptotically approaches infinity) this process leads to a scale-free network. The degree distribution of $G(p)$ satisfies a power law with the parameter for exponent being $\beta = 2 + \frac{p}{2-p}$. Scale-free networks also exhibit \emph{hierarchicity}. The local clustering coefficient is proportional to a power of the node degree
\begin{equation}\label{eqn:clustering-power-law}
C(k)\approx k^{-\alpha}
\end{equation}
where $\alpha$ is called the \emph{hierarchy coefficient}.

This distribution implies that the low-degree nodes belong to very dense sub-graphs and those sub-graphs are connected to each other through hubs. In other words, it means that the level of clustering is much larger than that in random networks.

Consequently, many of the network properties in a scale-free network are determined by the local structures -- namely, by a relatively small number of highly connected nodes (hubs). A consequence of this structure of the scale-free network is its extreme robustness to failure, a property also displayed by biomolecular networks and their modular structures. Such networks are highly tolerant of random failures (perturbations); however, they remain extremely sensitive to targeted attacks.

\subsubsection*{Assortativity Network Model}

{\em Assortative mixing} refers to the property exhibited by a preference of nodes to attach to similar (respectively, dissimilar) nodes; for example, high-degree vertices exhibit preference to attach to high-degree (resp.~low-degree) vertices. Network models, discussed earlier and including the preferential attachment model, do not capture such important properties exhibited by real biomolecular networks \cite{MR1908073}. Assortativity can be measured by the Pearson correlation coefficient $r$ of degrees of linked nodes \cite{MR1908073}. Positive correlation means connections between nodes of similar degree (assortativity) and negative correlation means connections between nodes with different degree (disassortativity). Unlike technological networks and social networks (showing assortative mixing), biological networks appear to evolve in a disassortative manner.

Many genetic networks, especially the DNA networks, lead to directed graphs. Assortative mixing can be generalized to directed biological graphs \cite{MR2982456}. For directed networks two new measures, in-assortativity and the out-assortativity , can be defined measuring the correlation between the in-degree $r_{in}$ and out-degree $r_{out}$ of the nodes respectively. Biological networks, which have been previously classified as disassortative, have been shown to be assortative with respect to these new measures. Also it has been shown that in directed biological networks, out-degree mixing patterns contain the highest amount of Shannon information, suggesting that nodes with high local out-assortativity (regulators) dominate the connectivity of the network \cite{MR2982456}. The occurrence of assortativity in social networks has been attributed to a process of homophily (that is people tend to associate with others on the basis of ethnicity, religion, sports preferences etc \cite{homo, homo1}). The mechanisms that give rise to assortativity in biomolecular networks likely arises by a similar proximate mechanism of like nodes forming edges with like nodes, but the ultimate cause(s) remains unclear.


\subsubsection*{Duplication Model}\label{subsubsec:Duplication}

Our earlier discussions suggest that biomolecular networks exhibit power-law degree distribution. However, unlike other complex networks, such as the Internet,  the growth exponent of biomolecular networks typically falls into a lower range $1<\beta <2$, as opposed to $\beta \geq 2$. This discrepancy has been suggested to have resulted from evolution by gene duplication dominating evolutionary mechanism  \cite{pmid:14633392}. Various biomolecular networks have been studied using a partial duplication process, which proceeds in the following manner: Let the initial graph $G_{0}$ have $N_0$ vertices. In each step, $G_t$ is constructed from its previous graph $G_{t-1}$ as follows: A random vertex $u$ is selected. Then a new vertex $v$ is added in such a way that for each neighbor $w$ of $u$, a new edge $(u, w)$ is added with probability $p$. The process is then applied repeatedly. The full duplication model is simply the partial model with $p=1$.

It has been shown that as the number $N$ of vertices becomes infinitely large, the partial duplication model with selection probability $p$ generates power-law graphs with the exponent satisfying the transcendental
equation  \cite{pmid:14633392}
$$p(\beta-1)=1-p^{\beta-1},$$
whose solution determines the scale-free exponent $\beta$ as a function of $p$. In particular,  if $1/2 <p <1$ then $\beta < 2$.

For illustrative purposes, we describe below an abstract gene network growth model incorporating the processes of gene duplication and deletion, as described above ( \cite{Mishra-Zhou-2004} and \cite{Zhou05}). Using a Markov chain model the following features were investigated: (i) the origination of the  segmental duplication; (ii) the effect of the duplication on the genome structure; and (iii) the role of duplication and deletion process in the genomic evolutionary distance. Unlike standard models of stationary Markov chain models, most processes in evolutionary biology belong to the group of non-stationary Markov processes, in which the transition matrix changes over time, or depends upon the current state.


This model results in the neutral emergence of scale-free degree distributions. It shows that the genomes of different organisms exhibit different network properties, likely reflecting differences in the rates of gene duplication and deletion  \cite{Mishra-Zhou-2004}. This analysis provides an example of how network topology can be used to provide insight into fundamental molecular evolutionary (neutral/Markov) processes in different species. Note that the model is relatively idealized, as it does not account for higher order interactions in a population involving: effective population size and allelic fixations; sex, diploidy and sex-chromosomes (e.g., X and Y in mammals or W and Z in birds, etc.); surveillance and repair in somatic cells; embryonic lethality; homologous recombination, etc. The mathematical model explored here is kept simple to motivate the machinery from graph theory developed later.

{\subsubsection*{Hierarchical Network Models}\label{subsubsec:Hierarchical}


Another interesting model, introduced by Ravasz and Barabasi  and dubbed {\it hierarchical network model},  simulates the characteristics of  many real life complex models and may be relevant.  The resulting networks have  modularity, high degree of clustering, and scale-free property. Modularity refers to the network phenomenon where many sparsely inter-connected dense subgraphs can be identified -- ``one can easily identify groups of nodes that are highly interconnected with each other, but have only a few or no links to nodes outside of the group to which they belong to.'' (from  \cite{ravasz03-1}).}

A generative process for hierarchical network model may be described as follows: For instance, consider an initial network $H_0$ of $c$ fully interconnected nodes (e.g., $c=5$). As a next step, create $(c-1)$ replicas of this cluster $H_0$ and connect the peripheral nodes of each replica to the central node of the original cluster to create $H_1$ with $c^2$ (e.g., $c^2 = 25$ ) nodes. This step can be repeated recursively and indefinitely, thereby for any $k$ steps the number of nodes generating the graph $H_k$ with $c^{k+1}$ nodes. If the central nodes of $H_0$ is called a \emph{hub} and other nodes \emph{peripheral}, then each recursion replicates additional copies of hubs and peripheral nodes.


One can carry out carry out a recursive analysis and shows that one obtains a power-law (i.e. scale-free) network with exponent
$\beta= 1 + \frac{\ln(c)} {\ln{(c-1)} }.$ The local clustering coefficients (for the hub-nodes)  follow
$C(k)\approx \frac{2}{k}$.
Also, one can show that
this  duplication feature of evolution leads to hierarchical behavior of the network. The networks are expected to be fundamentally modular, in other words, the network can be seamlessly partitioned into collection of modules where each module performs an identifiable task, separate from the function(s) of other  modules.
One can also show that the average clustering coefficient on $N$ nodes at any given stage is about $C=.7419282..$ (for $c=4$), $C=0.741840$ (for $c=5$), and a constant for a fixed $c$,  independent of $N$ (see \cite{ravasz03-1}, and for exact computations  \cite{PhysRevE.67.045103}). While for the preferential attachment model of Barbasi-Albert has the average clustering coefficient $C$ on $N$ nodes decreases as $1/N$, in addition not exhibiting modularity.

 




\section{Open Problems and Future Challenges}\label{sec:Open Problems}

The study of biomolecular networks is still a relatively young field and has thus far focused on a mechanistic perspective. As we begin to explore it from an evolutionary view point, we encounter a large array of promising areas of investigation -- most of which focuses on how information asymmetries among the gene players ultimately sculpt the information flow, as necessary for an organism to navigate in a complex and fluctuating environment. In particular, at its core this program requires an explanation of how features of genome evolution and structure might be algorithmically inferred from a network science perspective.

The traditional approaches of phylogenetic study may be applied here, but examining specifically the family of species-specific biomolecular networks. Thus mathematically we would need the networks to be aligned, motifs to be mapped to each other and network-distances to be correlated to deep evolutionary time. In order to account for the evolution by duplications, {\em orthologs} and {\em paralogs} of a gene (or gene families) are to be identified and connected to their roles in biochemical pathways. Ultimately, this analysis could be targeted at extracting the origin of various information-asymmetric signaling games and how they stabilized in their Nash equilibria.

Network analysis is used in disease studies,
but there have been more focused studies 
with applications to disease processes
in cancer. In Figure \ref{fig:Interactome} we show part of an interactome network useful in deciphering aberrant interactions in diseases (Figure 2.3 from \cite{JLABES-book2017}).

\subsection{Algorithmic Complexity Issues}\label{sec:Algorithmic Complexity}

A key problem central to this program would be in detecting isomorphism mappings among pairs of graphs or subgraphs, a problem of infeasible algorithmic complexity (assuming $P\neq NP$.) We start with a discussion of these issues and  cite heuristics that can tame the problem, albeit computing the solutions approximately. 

\subsubsection*{Intractability: NP-Completeness}

Many combinatorial optimization problems seem impossible to solve except by brute-force searches evaluating all possible configurations in the search space. They belong to a complexity class called NP-complete and include such problems as whether a graph has a clique of size $k$. Since finding certain shared motifs in a class of networks shares many computational characteristics of the clique problem and since it could be central to discovering important evolutionary signatures (e.g., EBD), it seems unlikely that it would be possible to characterize the evolutionary trajectories precisely -- especially when the number of genes involved are in the thousands.  
See the supplement for additional discussions on graph representations and to derive their algebraic invariants, that provide bounds on  complexity of algorithms possibly leading to  excellent  approximate results in the study of sparse complex networks (see \cite{MR1421568, MR2248695}. 

\paragraph*{Problem 4.A} {\em Classify various computational problems involved in detecting evolutionary trajectories of biomolecular networks and characterize their algorithmic complexity.\/}

\paragraph*{Problem 4.B} {\em Explore PTAS (Polynomial Time Approximation Schemes) for these problems -- Especially when the graphs satisfy certain sparsity, modularity and/or hierarchy properties.\/}

\subsubsection*{Algebraic Approximation}

As described earlier, many interesting topological features of a graph can be computed efficiently (on both sequential and parallel computers) from their descriptions in terms of adjacency matrices. The resulting spectral methods  have found recent applications  in complex networks (e.g., communication, social, Internet) (see \cite{Spielman-LectNotes}, \cite{MR1465732}, \cite{Spielman-Teng2014}, \cite{Spielman-Teng}, \cite{Spielman-Teng-2011}, \cite{Spielman-Teng-2004} \cite{MR2248695}, \cite{MR1421568}, \cite{MR2674816}, \cite{MR2012999}). These methods are efficient (linear time complexity) for sparse graphs, whose number of edges is roughly of the same order as the number of vertices. Thus, they are well suited to biomolecular networks (for example for clustering, community detection, hubs,  robustness, assortative mixing,  spreading and mixing, closeness, isomorphism,  among others).

Thus, spectral graph theory may be expected to have many applications in the analysis of biomolecular networks, most prominently, in clustering, graph similarity and graph approximation, but also in smoothing analysis and sparsification. 
One can envisage that many, if not most, classical network algorithms in biomolecular networks can be made faster by spectral methods. Indeed, since most biomolecular networks are sparse -- both in terms of sparse connections, and in precise algebraic sense (see the supplementary section), these algorithms likely lead to linear time algorithms. The smoothing analysis methods, as well as sparsification approximations are worth exploring in these contexts.

Another fruitful direction is in parallelizing these algorithms. As an illustration,  in several studies of biomolecular networks  it would be useful to identify when two networks $X_1$ and $X_2$ are ``close.'' We may wish to say that two networks are close if $Spec(X_1)$ and $Spec(X_2)$ are close -- a computational problem that is polynomially computable (and efficiently parallelizable) (see \cite{Spielman-Teng}). We can now give a mathematical formulation of this closeness, which can also be incorporated into phylogenetic studies.
These biomolecular networks may be annotated with weights that are linear or quadratic approximation of relations, as common in these studies. These analyses may identify subnetworks that have been influenced by EBD, in concert with selection.

\paragraph*{Problem 4.C} {\em Classify various algebraic problems involved in detecting evolutionary trajectories of biomolecular networks and characterize their ability to approximate. Explore their practical implementations on sequential and parallel computers.\/}

\subsection{Design Principles via Motif Analysis}

The study of Systems Biology postulates that there are important design principles of biological circuits that provide a great deal of insight. The connections of gene and protein interaction networks are assumed to provide the necessary robustness and control to achieve cellular function in the face of chemical noise. However, it remains unclear how random variations alone provide such robustness. A possible explanation may come from a game-theoretic model that lead to stable equilibria and is  expected to have precipitated from duplication of genes, interactions and motifs.  

\subsubsection*{Machine Learning}

The biomolecular networks of interest are derived from highly noisy data e.g., CHIP-Chip, CHIP-Seq (for GRN) or co-localization or two-hybrid (for PPI) and consequently, the inferred edges of the network may miss certain genuine interactions or include several spurious interactions. Various machine learning algorithms (with fdr, false discovery rates, control and regularization techniques) have been devised in order to improve the accuracy of such models. Biomolecular networks from related species (with orthologs and paralogs analysis) are often combined to improve the accuracies and cross-validate results. The accuracies may be further ascertained via various local properties. 

One important local property of networks are so-called network motifs, which are defined as recurrent and statistically significant sub-graphs or patterns. Thus, network motifs are sub-graphs that repeat themselves in a specific network or even among various networks. Each of these sub-graphs, defined by a particular pattern of interactions between vertices, may reflect a framework in which particular functions are achieved efficiently. Indeed, motifs are of notable importance largely because they may reflect functional properties. They have recently gathered much attention as a useful concept to uncover structural design principles of complex networks. Although network motifs may provide a deep insight into the network’s functional abilities, their detection is computationally challenging. Thus an important challenge for both experimental and computational scientists would be to study the evolutionary dynamics starting with the experimental data {\em ab initio}, as well as in improving the accuracy and efficiency of both the experimental and algorithmic techniques simultaneously.

\paragraph*{Problem 4.D} {\em Classify the species distributions of the different forms of heavy tailed distributions (e.g. power law, exponential, power law with exponential decay, lognormal), in different types of biomolecular network, and infer the mechanistic causes during network growth, and ultimate molecular evolutionary origins}

\paragraph*{Problem 4.E} {\em Characterize the motifs in the biomolecular networks of closely related species starting with the noisy experimental data. Explain the structure of the motifs via their effect on the information flow. For instance, one may focus on DOR (Dense Overlapping Regulons) motifs and how they might have evolved from a simpler ancestral regulon \cite{UriAlon}.}


\paragraph*{Problem 4.F} {\em Study Subgraph Isomorphism Algorithms (and heuristics) for sparse graphs and identify special cases most suitable for studying evolutionary trajectories, while relating them to biomolecular design principles.}

\subsubsection*{Network Alignment}

Critical to the evolutionary studies, described above, is the topic of network alignment and subsequent network tree building. Networks may be aligned in a pairwise fashion to calculate similarity, and from this a distance matrix calculated, and used for the construction of a network tree, showing the relationships between multiple networks. For example, in the case of meta-metabolic networks, such studies will reveal relationships between the meta-metabolic networks of different microhabitats. A plausible prediction is that the network tree should show convergent evolution in microbial communities from microhabitats with similar conditions (e.g., anaerobic habitats). Thus this approach could lead to a tool to study convergent evolution of microbial community structure in similar habitats \cite{emerge}.

From an algorithmic point of view, one may employ any of the three types of network alignment approaches:
\begin{enumerate}
\item where node identity is known;
\item where node similarity can be determined (based on sequence similarity for example); and
\item where node identity is unknown, here only network topology is used for alignment.
\end{enumerate}

\begin{figure}[ht]
\begin{center}
\vspace{130pt}%
\hspace*{-4cm}
\includegraphics[trim = 1cm 2cm 4cm 6cm, clip =false, width=0.6\textwidth]{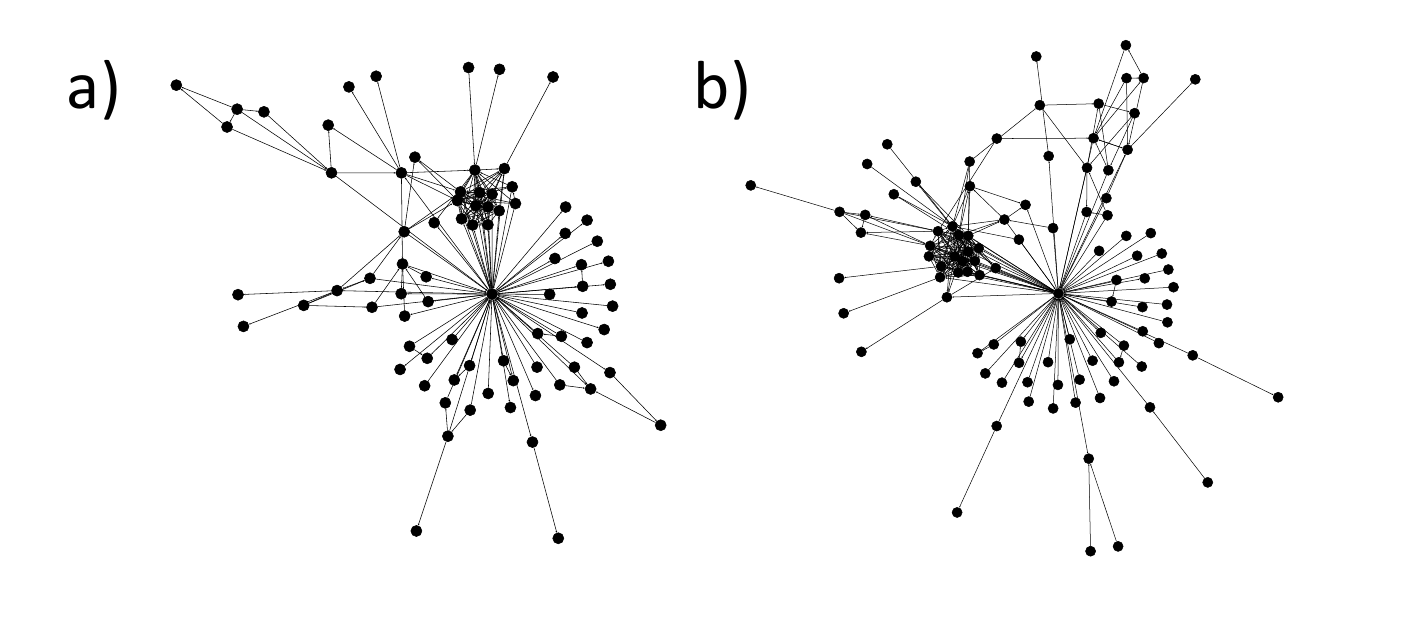}
\end{center}
\vspace{40pt}%
\caption{{\em Topological Alignment of Networks.\/} Similar Biomolecular networks could be topologically aligned and compared in order to express an evolutionary distance, which may then augment the traditional approaches of phylogenetic study. In order to account for the evolution by gene duplications, genes (or gene families) are to be identified and connected to their roles in biochemical pathways. Such an approach would lead to a program to understand the critical role of information asymmetries in driving evolution. Network alignment, a core problem in this program, is computationally intractable. To sharpen our intuition, we illustrate the problem using the social networks of the Gospels of Matthew and Luke. These networks represent social interactions between characters in the gospels of Matthew (a) and Luke (b). These were chosen as a basic test for topological alignment procedures, given that they share a similar number of nodes, and the highly connected node of Jesus. A straightforward test for the efficacy of a topological alignment algorithm therefore constitutes aligning both networks and verifying that the Jesus node from both networks is matched} \label{fig:3}
\end{figure}

\begin{figure}[ht]
\begin{center}
\includegraphics[width=4in, height=4in]{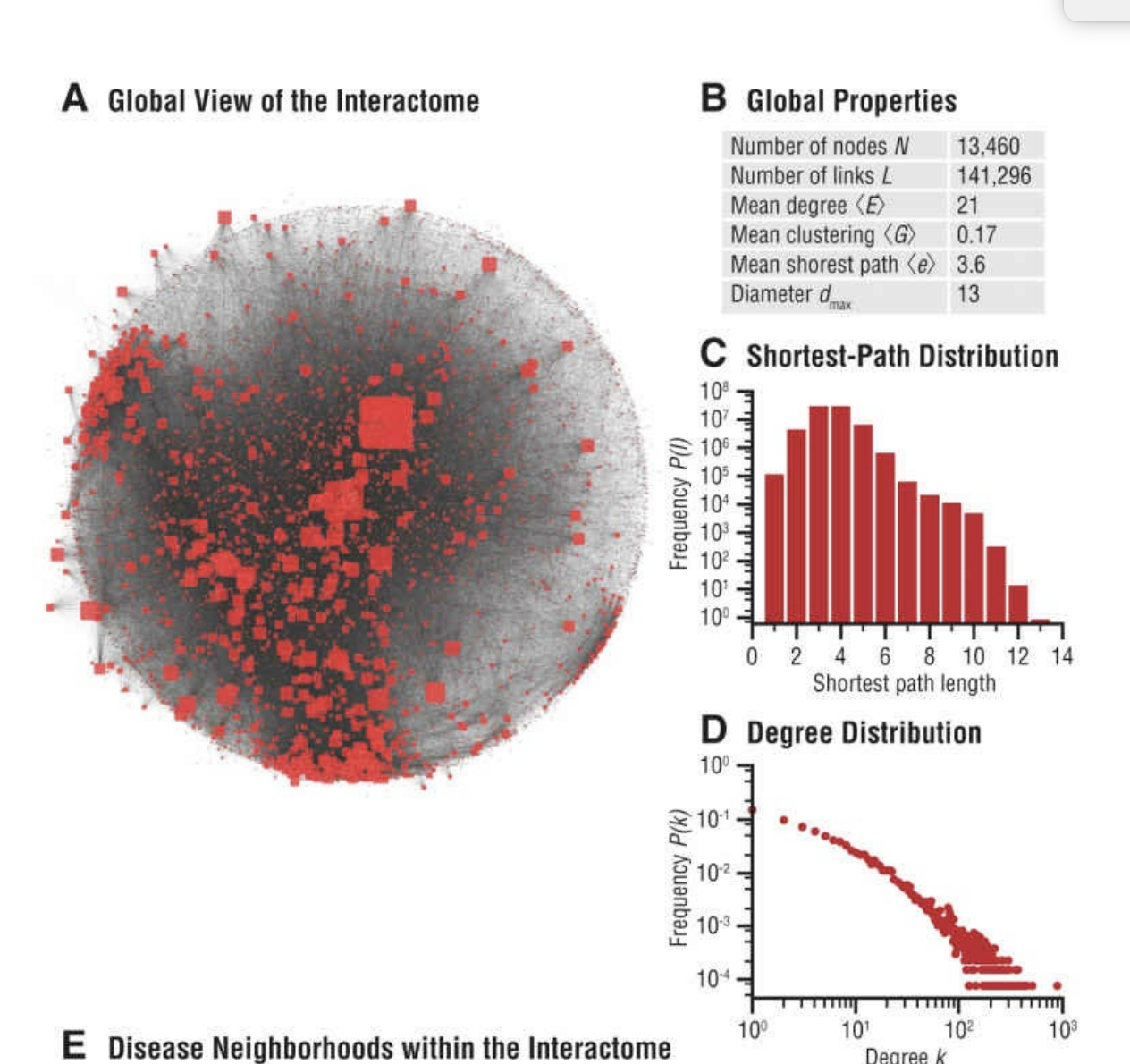}
\end{center}
\caption{
{\em Interactome Networks Used in the study of Diseases.} Undesirable interactions within a biomolecular network result in various disease states. Disease neighborhoods within the interactome can then be mapped to understand the progression of the disease. Progression of cancer have been studied using analysis of functionalization of oncogenes and dysfunctionalization of tumor suppression genes via copy number fluctuations, but much more can be learned from the topological features of these genes in their interaction neighborhood.
 (A) Global map of the interactome,
illustrating its heterogeneity. Node sizes are proportional to their degree, that is, the number of links each node has to
other nodes. (B) Basic characteristics of the interactome. (C) Distribution of the shortest paths within the interactome. The
average shortest path is $\langle d\rangle = 3.6.$ (D) The degree distribution of the interactome is approximately scale-free." (from  Figure 2.3 in  
\cite{JLABES-book2017}) }\label{fig:Interactome}
\end{figure}

The first is a straightforward edge alignment. However, a refined expression is required that incorporates similarities in edge widths in addition to the basic edge alignment (presence / absence of common edges between networks). There do exist some first generation heuristics that utilize the second and third types of alignment approach (i.e., sequence similarity and topology, and only topology) \cite{migraal}, but the underlying graph isomorphism problem is known to be $\#$P-complete. But these heuristics, as would be expected, do not work well -- a straightforward test for this problem is applying them to align the social networks of the Gospels of Luke and Matthew (Figure \ref{fig:3}) - the Jesus node should always align, as it is rather obvious topologically; but often leads to failure.

\paragraph*{Problem 4.G} {\em Classify and characterize the graph alignment algorithms.}

\subsection{Somatic Evolution and Cancer}

Cancer is a complex disease, but governed by somatic genomic evolution, as propelled by mutation. Thus as a consequence GRNs may be used to better understand cancer susceptibility, map its progression, design better tailored therapies, and better understand the evolution of endogenous anti-cancer strategies. Cancer genes are often network hubs \cite{hubs}, as they are often involved in critical developmental pathways. But a better network analysis will shed light on many natural questions: Why is it so? How does this come about from the process of network growth over evolutionary time? What clues do they provide to understand the somatic evolution in cancer and its progression?

During cancer progression, the disease reduces a cell's healthy genome into an aberrant mutant, where cancer eventually leads to metastasis, ultimately resulting in death of the patient. The healthy cells in the patient may be thought to possess a normal network, that is a gene network that engenders health and well-being. Cancer progression is reflected by a dynamic change of the normal network into an aberrant network. The aberrant network manifests itself by tumorigenesis, and finally metastasis.  There is a substantial literature enumerating the identity of oncogenes and tumor suppressor genes, which aberrantly gain function (e.g., amplification of copy number) or lose function (e.g., deletion in copy number, hemi- or homozygously), respectively. They modify the cell biology of cancer progression, effected via the dynamics of GRN and PPI networks in cancer progression -- all remain to be fully characterized. 

Of particular interest is the question whether there is an identifiable phase transition in network topology associated with metastasis. Figure 2 shows a simple model for how the evolution of p53 and its paralogs may affect GRN topology; such molecular evolutionary signaling games approaches may help to better understand the motifs associated with oncogenes in GRNs. An additional important factor in cancer is the pervasive occurrence of molecular deception \cite{escape}, which from a signaling games perspective is consistent with cancer's conflict of interest with somatic cells. The identity of deceptive macromolecular signals may be incorporated into the network, potentially shedding a novel light on the mechanism of carcinogenesis. The genesis of deceptive signals therefore is expected to impact and drive carcinogenesis.

An additional factor to understanding this biology are copy number variants (CNVs) -- types of gene mutations where a number of large sections of genomic DNA may be duplicated (or deleted), resulting in dosage effects of the resident gene sequences, which are exactly duplicated (or deleted). The numbers of CNVs can commonly vary substantially within a population, and have been shown to have significant roles in the propensity to develop cancer \cite{cnv}. An increase in the number of CNVs would have the effect of enhancing the weight of an edge, which represents the interaction of the CNV gene product with its macromolecular binding partner. Such a network variant represents an increased disposition to develop cancer, and can be understood as occupying a position in  `network space' (the space of all possible network topologies) in greater proximity to an aberrant network, than a normal network.

\paragraph*{Problem 4.H} {\em Study Cancer progression models in terms of GRN's and identify the role of {\em driver} and {\em passenger} genes in the somatically evolving networks.}

\subsection{Gene Regulation and 3D Networks}

In the genome of the ancestral life form, once a number of genes with separate function had evolved, it then would have become beneficial to evolve gene regulation. Therefore, genes with the dedicated function of regulating other genes in the genome would have arisen (transcription factors). The combination of regulatory and functional genes would have comprised the first gene regulation network. Increases in organismal complexity have been facilitated by an increase in the complexity of the gene regulation network \cite{burt2}.

Recent work has outlined the importance of three-dimensional proximity of genes to genes on other chromosomes, in addition to their immediate neighborhood on their own chromosome \cite{3D}. This effect implies that gene proximity and spatial relationships within the nucleus can be meaningfully represented as a network. Such a network would be comprised of two types of edge: 1) linear distance on the same chromosome (centimorgans), 2) physical distance with genes on other chromosomes (nanometers). Such networks may be termed 3D gene orientation networks.

Gene regulation and co-regulation may be better understood by the construction and analysis of 3D gene orientation networks. This is because the proximity of regulatory modules to a gene has an influence of gene expression. Most genes have a regulatory region 5' of the transcription start site, the promoter. In addition, regulatory enhancers and other regulatory elements may be located distant from the gene, generally on the same chromosome \cite{enhance}. It is thought that the bending and juxtaposition of chromosomes within the nucleus may bring such elements into physical proximity to the gene \cite{enhance}. Clearly, the physical distance, and frequency with which the element is brought into contact with the gene will influence the nature of its regulatory input. Using 3D gene orientation networks, additional information may be incorporated into edges, such as whether physical proximity is static, or has movement. If there is movement, this may be coordinated (or not) with other regulatory elements affecting the same gene. Likewise, interactions with regulatory elements may show some coordination between genes.

\paragraph*{Problem 4.I} {\em Describe the Gene Duplication process and their utilities in terms of the genome's 3D structure.}

\section{Conclusion}
Here, we have outlined graph theoretical approaches that may reveal some novel aspects of the molecular evolutionary process, which become manifest at the level of the phenome. Further work is required to link the diverse features of network topology with network evolution and growth. While the evolutionary aspects shaping individual gene-gene interactions has been addressed by geneticists and molecular evolutionists, we believe that a multi-disciplinary effort combining game theory, graph theory, and algebraic/statistical analysis will provide a more informative omnigenic model of gene interactions, in contrast to the traditional homogenic view. Given our view that biomolecular networks may be modeled using evolutionary game theory, and game theoretical approaches in the study of social networks, we expect that some surprising similarities and convergences between the topologies of the two might be observed. Finally, we note that the field of statistics gained impetus from the consideration of biological problems, from workers such as Fisher, Haldane, Rao, Wright, Kimura, Crow and others, and so we suggest that consideration of the open problems listed here might also lead to a similar development of new mathematics.

\section{Bibliographic Notes}
We recommend the following articles for further reading: (\cite {MR3045557}, \cite {MR2643139}, \cite {MR2608985}, \cite {MR2457138}, \cite {MR2453864}, \cite {MR2443946}, \cite{MR2108989}, \cite{MR2150394}, \cite{MR2016119}, \cite {MR1961687}, \cite {MR1943379}, \cite {MR1919733},  \cite {MR1895096}), \cite{MR2108973},
\cite{MR2248695} and  \cite{MR37}. For other important sources (especially with respect to directed graphs), we refer to \cite {MR3713556},\cite {MR3678034},
(\cite {MR2736374}, \cite {MR2676073}, \cite {MR2563829}, \cite {MR2282142}, \cite {MR2282139}, \cite {MR2250459},\cite {MR2108985},\cite {MR2016118},\cite {MR2010377},\cite {MR1975193}, \cite {MR1908073},\cite {MR1812610},\cite {MR1732095}), \cite{newman2011structure}. For network alignments and evolution of networks see for example \cite{Sharan1974, Pinter2005, Kalaev2008, Mazurie2010} . For bipartite networks (\cite{MR2988185} and  \cite{MR1959169}). For Spectral methods (\cite{MR572262},
\cite{MR1421568},\cite{Spielman-Teng-2011},
\cite{MR2825307},\cite{MR2248695},\cite{MR1308046},
\cite{MR37},\cite{MR963118},\cite{Sarnak:AMS},
\cite{MR1989434}, and \cite{MR2869010}).

\section*{Author Contributions}

B.M.~conceived of and structured the presented ideas at a high level. S.M. and B.M.~developed the biological theories and H.J.~\& J.V.~developed the computational, quantitative and mathematical theories.
All authors discussed the open problems and contributed to the final manuscript.

\section*{Funding}

This work was supported by National Science Foundation Grants CCF-0836649 and CCF-0926166, and a National Cancer Institute Physical Sciences-Oncology Center Grant U54 CA193313-01 (to B.M.).

\section*{Acknowledgments}

We acknowledge our colleagues in UPR and NYU, who have generously provided many constructive criticisms.


\section*{Supplemental Data}
 \href{http://home.frontiersin.org/about/author-guidelines#SupplementaryMaterial}{Supplementary Material} should be uploaded separately on submission, if there are Supplementary Figures, please include the caption in the same file as the figure. LaTeX Supplementary Material templates can be found in the Frontiers LaTeX folder.

\section*{Data Availability Statement}
The datasets [GENERATED/ANALYZED] for this study can be found in the [NAME OF REPOSITORY] [LINK].

\bibliographystyle{frontiersinSCNS_ENG_HUMS}
\bibliography{bibliography}

\begin{thebibliography}{119}
\providecommand{\natexlab}[1]{#1}
\expandafter\ifx\csname urlstyle\endcsname\relax
  \providecommand{\doi}[1]{doi:\discretionary{}{}{}#1}\else
  \providecommand{\doi}{doi:\discretionary{}{}{}\begingroup
  \urlstyle{rm}\Url}\fi
\providecommand{\selectlanguage}[1]{\relax}
\providecommand{\bibAnnoteFile}[1]{%
  \IfFileExists{#1}{\begin{quotation}\noindent\textsc{Key:} #1\\
  \textsc{Annotation:}\ \input{#1}\end{quotation}}{}}
\providecommand{\bibAnnote}[2]{%
  \begin{quotation}\noindent\textsc{Key:} #1\\
  \textsc{Annotation:}\ #2\end{quotation}}

\bibitem[{Albert and Barab\'{a}si(2002)}]{MR1895096}
Albert, R. and Barab\'{a}si, A.-L. (2002).
\newblock Statistical mechanics of complex networks.
\newblock \emph{Rev. Modern Phys.} 74, 47--97.
\newblock \doi{10.1103/RevModPhys.74.47}
\bibAnnoteFile{MR1895096}

\bibitem[{Alexander et~al.(2012)Alexander, Skyrms, and Zabell}]{skyrm}
Alexander, J., Skyrms, B., and Zabell, S. (2012).
\newblock Inventing new signals.
\newblock \emph{Dynamic Games and Applications} 2, 129--145
\bibAnnoteFile{skyrm}

\bibitem[{Alon(2006)}]{UriAlon}
Alon, U. (2006).
\newblock \emph{An introduction to systems biology: design principles of
  biological circuits} (Chapman and Hall/CRC)
\bibAnnoteFile{UriAlon}

\bibitem[{Barab\'{a}si(2003)}]{MR2016119}
Barab\'{a}si, A.-L. (2003).
\newblock Emergence of scaling in complex networks.
\newblock In \emph{Handbook of graphs and networks} (Wiley-VCH, Weinheim).
  69--84
\bibAnnoteFile{MR2016119}

\bibitem[{Barab\'{a}si(2009)}]{MR2548299}
Barab\'{a}si, A.-L. (2009).
\newblock Scale-free networks: a decade and beyond.
\newblock \emph{Science} 325, 412--413.
\newblock \doi{10.1126/science.1173299}
\bibAnnoteFile{MR2548299}

\bibitem[{Barab\'{a}si and Albert(1999)}]{MR2091634}
Barab\'{a}si, A.-L. and Albert, R. (1999).
\newblock Emergence of scaling in random networks.
\newblock \emph{Science} 286, 509--512.
\newblock \doi{10.1126/science.286.5439.509}
\bibAnnoteFile{MR2091634}

\bibitem[{Barab\'{a}si et~al.(2003)Barab\'{a}si, Dezs\H{o}, Ravasz, Yook, and
  Oltvai}]{MR2150394}
Barab\'{a}si, A.-L., Dezs\H{o}, Z., Ravasz, E., Yook, S.-H., and Oltvai, Z.
  (2003).
\newblock Scale-free and hierarchical structures in complex networks.
\newblock In \emph{Modeling of complex systems} (Amer. Inst. Phys., Melville,
  NY), vol. 661 of \emph{AIP Conf. Proc.} 1--16.
\newblock \doi{10.1063/1.1571285}
\bibAnnoteFile{MR2150394}

\bibitem[{Barab\'{a}si et~al.(2002)Barab\'{a}si, Jeong, N\'{e}da, Ravasz,
  Schubert, and Vicsek}]{MR1943379}
Barab\'{a}si, A.~L., Jeong, H., N\'{e}da, Z., Ravasz, E., Schubert, A., and
  Vicsek, T. (2002).
\newblock Evolution of the social network of scientific collaborations.
\newblock \emph{Phys. A} 311, 590--614.
\newblock \doi{10.1016/S0378-4371(02)00736-7}
\bibAnnoteFile{MR1943379}

\bibitem[{Barab\'{a}si et~al.(2004)Barab\'{a}si, Oltvai, and
  Wuchty}]{MR2108989}
Barab\'{a}si, A.-L., Oltvai, Z.~N., and Wuchty, S. (2004).
\newblock Characteristics of biological networks.
\newblock In \emph{Complex networks} (Springer, Berlin), vol. 650 of
  \emph{Lecture Notes in Phys.} 443--457.
\newblock \doi{10.1007/978-3-540-44485-5_20}
\bibAnnoteFile{MR2108989}

\bibitem[{Belyi et~al.(2010)Belyi, Ak, Markert, Wang, Hu, and Levine}]{p53}
Belyi, V., Ak, P., Markert, E., Wang, H., Hu, A., W. Puzio-Kuter, and Levine,
  A. (2010).
\newblock The origins and evolution of the p53 family of genes.
\newblock \emph{Cold Spring Harbor Perspect Biol} 2, a001198
\bibAnnoteFile{p53}

\bibitem[{Bhatia and Kumar(2013)}]{escape}
Bhatia, A. and Kumar, Y. (2013).
\newblock Cellular and molecular mechanisms in cancer immune escape: A
  comprehensive review.
\newblock \emph{Expert Rev Clin Immunol} 10, 758--762
\bibAnnoteFile{escape}

\bibitem[{Biggs(1993)}]{biggs}
Biggs, N. (1993).
\newblock \emph{Algebraic graph theory}.
\newblock Cambridge Mathematical Library (Cambridge: Cambridge University
  Press), second edn.
\bibAnnoteFile{biggs}

\bibitem[{Burt and Trivers(2006)}]{burt}
Burt, A. and Trivers, R. (2006).
\newblock \emph{Genes in Conflict: The Biology of Selfish Genetic Elements}
  (Cambridge, Massachusetts: Harvard University Press)
\bibAnnoteFile{burt}

\bibitem[{Burton(2014)}]{burt2}
Burton, Z. (2014).
\newblock The old and new testaments of gene regulation.
\newblock \emph{Transcription} 5, e28674
\bibAnnoteFile{burt2}

\bibitem[{Candia et~al.(2008)Candia, Gonz\'{a}lez, Wang, Schoenharl, Madey, and
  Barab\'{a}si}]{MR2453864}
Candia, J., Gonz\'{a}lez, M.~C., Wang, P., Schoenharl, T., Madey, G., and
  Barab\'{a}si, A.-L. (2008).
\newblock Uncovering individual and collective human dynamics from mobile phone
  records.
\newblock \emph{J. Phys. A} 41, 224015, 11.
\newblock \doi{10.1088/1751-8113/41/22/224015}
\bibAnnoteFile{MR2453864}

\bibitem[{Chang et~al.(2017)Chang, Pannunzio, Adachi, and Lieber}]{nhej}
Chang, H., Pannunzio, N., Adachi, N., and Lieber, M. (2017).
\newblock Non-homologous dna end joining and alternative pathways to
  double-strand break repair.
\newblock \emph{Nature Reviews Molecular Cellular Biology} 18, 495--506
\bibAnnoteFile{nhej}

\bibitem[{Chung(2010)}]{MR2674816}
Chung, F. (2010).
\newblock Graph theory in the information age.
\newblock \emph{Notices Amer. Math. Soc.} 57, 726--732
\bibAnnoteFile{MR2674816}

\bibitem[{Chung and Lu(2004)}]{MR2108973}
Chung, F. and Lu, L. (2004).
\newblock The small world phenomenon in hybrid power law graphs.
\newblock In \emph{Complex networks} (Springer, Berlin), vol. 650 of
  \emph{Lecture Notes in Phys.} 89--104.
\newblock \doi{10.1007/978-3-540-44485-5_4}
\bibAnnoteFile{MR2108973}

\bibitem[{Chung and Lu(2006)}]{MR2248695}
Chung, F. and Lu, L. (2006).
\newblock \emph{Complex graphs and networks}, vol. 107 of \emph{CBMS Regional
  Conference Series in Mathematics} (Published for the Conference Board of the
  Mathematical Sciences, Washington, DC; by the American Mathematical Society,
  Providence, RI).
\newblock \doi{10.1090/cbms/107}
\bibAnnoteFile{MR2248695}

\bibitem[{Chung et~al.(2003)Chung, Lu, Dewey, and Galas}]{pmid:14633392}
Chung, F., Lu, L., Dewey, T.~G., and Galas, D.~J. (2003).
\newblock Duplication models for biological networks.
\newblock \emph{Journal of computational biology : a journal of computational
  molecular cell biology} 10, 677--87
\bibAnnoteFile{pmid:14633392}

\bibitem[{Chung(1997)}]{MR1421568}
Chung, F. R.~K. (1997).
\newblock \emph{Spectral graph theory}, vol.~92 of \emph{CBMS Regional
  Conference Series in Mathematics} (Published for the Conference Board of the
  Mathematical Sciences, Washington, DC; by the American Mathematical Society,
  Providence, RI)
\bibAnnoteFile{MR1421568}

\bibitem[{Clauset et~al.(2009{\natexlab{a}})Clauset, Shalizi, and
  Newman}]{clau}
Clauset, A., Shalizi, C., and Newman, M. (2009{\natexlab{a}}).
\newblock Power-law distributions in empirical data.
\newblock \emph{SIAM Rev} 51, 661--703
\bibAnnoteFile{clau}

\bibitem[{Clauset et~al.(2009{\natexlab{b}})Clauset, Shalizi, and
  Newman}]{MR2563829}
Clauset, A., Shalizi, C.~R., and Newman, M. E.~J. (2009{\natexlab{b}}).
\newblock Power-law distributions in empirical data.
\newblock \emph{SIAM Rev.} 51, 661--703.
\newblock \doi{10.1137/070710111}
\bibAnnoteFile{MR2563829}

\bibitem[{Cotterell et~al.(2017)Cotterell, Vylomova, Khayrallah, Kirov, and
  Yarowsky}]{word}
Cotterell, R., Vylomova, E., Khayrallah, H., Kirov, C., and Yarowsky, D.
  (2017).
\newblock Paradigm completion for derivational morphology.
\newblock \emph{Proceedings of the 2017 Conference on Empirical Methods in
  Natural Language Processing} , 714--720
\bibAnnoteFile{word}

\bibitem[{Crawford and Sobel(1982)}]{cs1}
Crawford, V.~P. and Sobel, J. (1982).
\newblock Strategic information transmission.
\newblock \emph{Econometrica} 50, 1431--1451
\bibAnnoteFile{cs1}

\bibitem[{Cvetkovi{\'c} et~al.(1980)Cvetkovi{\'c}, Doob, and Sachs}]{MR572262}
Cvetkovi{\'c}, D.~M., Doob, M., and Sachs, H. (1980).
\newblock \emph{Spectra of graphs}, vol.~87 of \emph{Pure and Applied
  Mathematics} (New York: Academic Press Inc. [Harcourt Brace Jovanovich
  Publishers]).
\newblock Theory and application
\bibAnnoteFile{MR572262}

\bibitem[{Davidoff et~al.(2003)Davidoff, Sarnak, and Valette}]{MR1989434}
Davidoff, G., Sarnak, P., and Valette, A. (2003).
\newblock \emph{Elementary number theory, group theory, and {R}amanujan
  graphs}, vol.~55 of \emph{London Mathematical Society Student Texts}
  (Cambridge University Press, Cambridge).
\newblock \doi{10.1017/CBO9780511615825}
\bibAnnoteFile{MR1989434}

\bibitem[{Davis et~al.(2010)Davis, Chawla, Christakis, and
  Barab\'{a}si}]{MR2608985}
Davis, D.~A., Chawla, N.~V., Christakis, N.~A., and Barab\'{a}si, A.-L. (2010).
\newblock Time to {CARE}: a collaborative engine for practical disease
  prediction.
\newblock \emph{Data Min. Knowl. Discov.} 20, 388--415.
\newblock \doi{10.1007/s10618-009-0156-z}
\bibAnnoteFile{MR2608985}

\bibitem[{Demuth et~al.(2018)Demuth, De~Bie, Stajich, Cristianini, and
  Hahn}]{demuth}
Demuth, J., De~Bie, T., Stajich, J., Cristianini, N., and Hahn, M. (2018).
\newblock The evolution of mammalian gene families.
\newblock \emph{PLoS One} 1
\bibAnnoteFile{demuth}

\bibitem[{Dodds and Watts(2005)}]{MR2125834}
Dodds, P.~S. and Watts, D.~J. (2005).
\newblock A generalized model of social and biological contagion.
\newblock \emph{J. Theoret. Biol.} 232, 587--604.
\newblock \doi{10.1016/j.jtbi.2004.09.006}
\bibAnnoteFile{MR2125834}

\bibitem[{Dokholyan et~al.(2002)Dokholyan, Shakhnovich, and Shakhnovich}]{bang}
Dokholyan, N., Shakhnovich, B., and Shakhnovich, E. (2002).
\newblock Expanding protein universe and its origin in from the biological big
  bang.
\newblock \emph{Proc Natl Acad Sci USA} 99, 14132--14136
\bibAnnoteFile{bang}

\bibitem[{Erd{\"o}s and R{\'e}nyi(1959)}]{erdos}
Erd{\"o}s, P. and R{\'e}nyi, A. (1959).
\newblock On random graphs.
\newblock \emph{Publicationes Mathematicae} 6, 290--297
\bibAnnoteFile{erdos}

\bibitem[{Farkas et~al.(2002)Farkas, Der\'{e}nyi, Jeong, N\'{e}da, Oltvai,
  Ravasz et~al.}]{MR1961687}
Farkas, I., Der\'{e}nyi, I., Jeong, H., N\'{e}da, Z., Oltvai, Z.~N., Ravasz,
  E., et~al. (2002).
\newblock Networks in life: scaling properties and eigenvalue spectra.
\newblock \emph{Phys. A} 314, 25--34.
\newblock \doi{10.1016/S0378-4371(02)01181-0}.
\newblock Horizons in complex systems (Messina, 2001)
\bibAnnoteFile{MR1961687}

\bibitem[{Girvan and Newman(2002)}]{MR1908073}
Girvan, M. and Newman, M. E.~J. (2002).
\newblock Community structure in social and biological networks.
\newblock \emph{Proc. Natl. Acad. Sci. USA} 99, 7821--7826.
\newblock \doi{10.1073/pnas.122653799}
\bibAnnoteFile{MR1908073}

\bibitem[{Goh and Barab\'{a}si(2008)}]{MR2443946}
Goh, K.-I. and Barab\'{a}si, A.-L. (2008).
\newblock Burstiness and memory in complex systems.
\newblock \emph{Europhys. Lett. EPL} 81, Art. 48002, 5.
\newblock \doi{10.1209/0295-5075/81/48002}
\bibAnnoteFile{MR2443946}

\bibitem[{Goldford et~al.(2018)Goldford, Lu, Bajic, Estrela, Tikhonov,
  Sanchez-Gorostiaga et~al.}]{emerge}
Goldford, J., Lu, N., Bajic, D., Estrela, S., Tikhonov, M., Sanchez-Gorostiaga,
  A., et~al. (2018).
\newblock Emergent simplicity in microbial community assembly.
\newblock \emph{Science} 361, 1390--1396
\bibAnnoteFile{emerge}

\bibitem[{Gondor and Ohlsson(2018)}]{enhance}
Gondor, A. and Ohlsson, R. (2018).
\newblock Enhancer functions in three dimensions: beyond the flat world
  perspective.
\newblock \emph{F1000Research} 7, 681
\bibAnnoteFile{enhance}

\bibitem[{Govindarajan and Goldstein(1997)}]{neutprot}
Govindarajan, S. and Goldstein, R. (1997).
\newblock Evolution of model proteins on a foldability landscape.
\newblock \emph{Proteins} 29, 461--466
\bibAnnoteFile{neutprot}

\bibitem[{Hartwell et~al.(1999)Hartwell, Hopfield, Leibler, and
  Murray}]{pmid:10591225}
Hartwell, L.~H., Hopfield, J.~J., Leibler, S., and Murray, A.~W. (1999).
\newblock From molecular to modular cell biology.
\newblock \emph{Nature} 402, C47--52
\bibAnnoteFile{pmid:10591225}

\bibitem[{Hawking and Hertog(2018)}]{hawk}
Hawking, S. and Hertog, T. (2018).
\newblock A smooth exit from eternal inflation?
\newblock \emph{arXiv:1707.07702}
\bibAnnoteFile{hawk}

\bibitem[{H\o~holdt and Janwa(2012)}]{MR2988185}
H\o~holdt, T. and Janwa, H. (2012).
\newblock Eigenvalues and expansion of bipartite graphs.
\newblock \emph{Des. Codes Cryptogr.} 65, 259--273.
\newblock \doi{10.1007/s10623-011-9598-6}
\bibAnnoteFile{MR2988185}

\bibitem[{Huang et~al.(2017)Huang, Liao, and Wu}]{ppi}
Huang, L., Liao, L., and Wu, C. (2017).
\newblock Evolutionary analysis and interaction prediction for protein-protein
  interaction network in geometric space.
\newblock \emph{PLoS One} 12, e0183495
\bibAnnoteFile{ppi}

\bibitem[{Innan and Kondrashov(2010)}]{innan}
Innan, H. and Kondrashov, F. (2010).
\newblock The evolution of gene duplications: classifying and distinguishing
  between models.
\newblock \emph{Nat Rev Genet} 11, 97--108
\bibAnnoteFile{innan}

\bibitem[{Janwa and Lal(2003)}]{MR1959169}
Janwa, H. and Lal, A.~K. (2003).
\newblock On {T}anner codes: minimum distance and decoding.
\newblock \emph{Appl. Algebra Engrg. Comm. Comput.} 13, 335--347.
\newblock \doi{10.1007/s00200-003-0098-4}
\bibAnnoteFile{MR1959169}

\bibitem[{Janwa and Rangachari(2015)}]{MR37}
Janwa, H. and Rangachari, S. (2015).
\newblock Ramanujan graphs and their applications
\bibAnnoteFile{MR37}

\bibitem[{Joerger and Fersht(2006)}]{p53cancer}
Joerger, A. and Fersht, A. (2006).
\newblock The p53 pathway: origins, inactivation in cancer, and emerging
  therapeutic approaches.
\newblock \emph{Annual Review of Biochemistry} 85, 375--404
\bibAnnoteFile{p53cancer}

\bibitem[{Kalaev et~al.(2008)Kalaev, Smoot, Ideker, and Sharan}]{Kalaev2008}
Kalaev, M., Smoot, M., Ideker, T., and Sharan, R. (2008).
\newblock Networkblast: comparative analysis of protein networks 24, 594--596.
\newblock \doi{10.1093/bioinformatics/btm630}.
\newblock Exported from https://app.dimensions.ai on 2018/11/21
\bibAnnoteFile{Kalaev2008}

\bibitem[{Karimzadeh et~al.(2018)Karimzadeh, Jandaghi, Papadakis, Trainor,
  Gonzalez-Porta, Scelo et~al.}]{hubs}
Karimzadeh, M., Jandaghi, P., Papadakis, A., Trainor, S., Gonzalez-Porta, M.,
  Scelo, G., et~al. (2018).
\newblock Aberration hubs in protein interaction networks highlight actionable
  targets in cancer.
\newblock \emph{Oncotarget} 9, 25166--25180
\bibAnnoteFile{hubs}

\bibitem[{Karp(2011)}]{Karp-2011}
Karp, R.~M. (2011).
\newblock Heuristic algorithms in computational molecular biology.
\newblock \emph{J. Comput. System Sci.} 77, 122--128.
\newblock \doi{10.1016/j.jcss.2010.06.009}
\bibAnnoteFile{Karp-2011}

\bibitem[{Karrer and Newman(2010)}]{MR2736374}
Karrer, B. and Newman, M. E.~J. (2010).
\newblock Message passing approach for general epidemic models.
\newblock \emph{Phys. Rev. E (3)} 82, 016101, 9.
\newblock \doi{10.1103/PhysRevE.82.016101}
\bibAnnoteFile{MR2736374}

\bibitem[{K\'ep\`es(2007)}]{Kepes-book-2007}
K\'ep\`es, F.~e. (2007).
\newblock \emph{Biological Networks}.
\newblock Complex Systems and Interdisciplinary Science (World Scientific)
\bibAnnoteFile{Kepes-book-2007}

\bibitem[{Krepischi et~al.(2012)Krepischi, Pearson, and Rosenberg}]{cnv}
Krepischi, A., Pearson, P., and Rosenberg, C. (2012).
\newblock Germline copy number variations and cancer predisposition.
\newblock \emph{Future Oncology} 8, 681
\bibAnnoteFile{cnv}

\bibitem[{Kuchaiev and Przulj(2011)}]{migraal}
Kuchaiev, O. and Przulj, N. (2011).
\newblock Integrative network alignment reveals large regions of global network
  similarity in yeast and human.
\newblock \emph{Bioinformatics} 27, 1390--1396
\bibAnnoteFile{migraal}

\bibitem[{Lespinet et~al.(2002)Lespinet, Wolf, Koonin, and Aravind}]{lwka02}
Lespinet, O., Wolf, Y.~I., Koonin, E.~V., and Aravind, L. (2002).
\newblock The role of lineage-specific gene family expansion in the evolution
  of eukaryotes.
\newblock \emph{Genome Research} 12, 1048--1059
\bibAnnoteFile{lwka02}

\bibitem[{Lewis(1969)}]{l}
Lewis, D. (1969).
\newblock \emph{Convention: a philosophical study} (Cambridge: Harvard
  University Press)
\bibAnnoteFile{l}

\bibitem[{Li et~al.(2018)Li, Hu, and Shen}]{3D}
Li, Y., Hu, M., and Shen, Y. (2018).
\newblock Gene regulation in the 3{D} genome.
\newblock \emph{Human Mol Genet} 27, R228--233
\bibAnnoteFile{3D}

\bibitem[{Liu et~al.(2013)Liu, Slotine, and Barab\'{a}si}]{MR3045557}
Liu, Y.-Y., Slotine, J.-J., and Barab\'{a}si, A.-L. (2013).
\newblock Observability of complex systems.
\newblock \emph{Proc. Natl. Acad. Sci. USA} 110, 2460--2465.
\newblock \doi{10.1073/pnas.1215508110}
\bibAnnoteFile{MR3045557}

\bibitem[{Loscalzo and Barab\'{a}si(2016)}]{Barbasi-book2016}
Loscalzo, J. and Barab\'{a}si, A.-L. (2016).
\newblock \emph{Network Science} (Cambridge University Press, Cambridge), 1st
  edn.
\bibAnnoteFile{Barbasi-book2016}

\bibitem[{Loscalzo et~al.(2017)Loscalzo, Barab\'{a}si, and
  Silverman}]{JLABES-book2017}
Loscalzo, J., Barab\'{a}si, A.-L., and Silverman, E. K.~e. (2017).
\newblock \emph{Network Medicine: Complex Systems in Human Disease and
  Therapeutic} (Harvard University Press), 1st edn.
\bibAnnoteFile{JLABES-book2017}

\bibitem[{Lu et~al.(2009)Lu, Amatruda, and Abrams}]{dup}
Lu, W.-J., Amatruda, J., and Abrams, J. (2009).
\newblock p53 ancestry: gazing through an evolutionary lens.
\newblock \emph{Nature Reviews Cancer} 9, 758--762
\bibAnnoteFile{dup}

\bibitem[{Lubotzky(1994)}]{MR1308046}
Lubotzky, A. (1994).
\newblock \emph{Discrete groups, expanding graphs and invariant measures}, vol.
  125 of \emph{Progress in Mathematics} (Basel: Birkh\"auser Verlag).
\newblock With an appendix by Jonathan D. Rogawski
\bibAnnoteFile{MR1308046}

\bibitem[{Lubotzky(2012)}]{MR2869010}
Lubotzky, A. (2012).
\newblock Expander graphs in pure and applied mathematics.
\newblock \emph{Bull. Amer. Math. Soc. (N.S.)} 49, 113--162.
\newblock \doi{10.1090/S0273-0979-2011-01359-3}
\bibAnnoteFile{MR2869010}

\bibitem[{Lubotzky et~al.(1988)Lubotzky, Phillips, and Sarnak}]{MR963118}
Lubotzky, A., Phillips, R., and Sarnak, P. (1988).
\newblock Ramanujan graphs.
\newblock \emph{Combinatorica} 8, 261--277.
\newblock \doi{10.1007/BF02126799}
\bibAnnoteFile{MR963118}

\bibitem[{MacKay(2003)}]{MR2012999}
MacKay, D. J.~C. (2003).
\newblock \emph{Information theory, inference and learning algorithms}
  (Cambridge University Press, New York)
\bibAnnoteFile{MR2012999}

\bibitem[{Massey(2015)}]{nemerge}
Massey, S. (2015).
\newblock Genetic code evolution reveals the neutral emergence of mutational
  robustness, and information as an evolutionary constraint.
\newblock \emph{Life} 5, 1301--1332
\bibAnnoteFile{nemerge}

\bibitem[{Massey and Mishra(2018)}]{mishmas}
Massey, S. and Mishra, B. (2018).
\newblock Origin of biomolecular games: deception and molecular evolution.
\newblock \emph{J Royal Soc Interface} 15, 20180329
\bibAnnoteFile{mishmas}

\bibitem[{Mazurie et~al.(2010)Mazurie, Bonchev, Schwikowski, and
  Buck}]{Mazurie2010}
Mazurie, A., Bonchev, D., Schwikowski, B., and Buck, G.~A. (2010).
\newblock Evolution of metabolic network organization.
\newblock \emph{BMC Systems Biology} 4, 59.
\newblock \doi{10.1186/1752-0509-4-59}
\bibAnnoteFile{Mazurie2010}

\bibitem[{McCloskey et~al.(2013)McCloskey, Palsson, and Feist}]{metab}
McCloskey, D., Palsson, B., and Feist, A. (2013).
\newblock Basic and applied uses of genome-scale metabolic network
  reconstructions of escherichia coli.
\newblock \emph{Mol Syst Biol} 9, 661
\bibAnnoteFile{metab}

\bibitem[{McPherson et~al.(2001)McPherson, Smith-Lovin, and Cook}]{homo}
McPherson, M., Smith-Lovin, L., and Cook, J. (2001).
\newblock Birds of a feather: homophily in social networks.
\newblock \emph{Ann Rev Sociol} 27, 415--444
\bibAnnoteFile{homo}

\bibitem[{Meyers et~al.(2006)Meyers, Newman, and Pourbohloul}]{MR2250459}
Meyers, L.~A., Newman, M. E.~J., and Pourbohloul, B. (2006).
\newblock Predicting epidemics on directed contact networks.
\newblock \emph{J. Theoret. Biol.} 240, 400--418.
\newblock \doi{10.1016/j.jtbi.2005.10.004}
\bibAnnoteFile{MR2250459}

\bibitem[{Mishra and Zhou(2004)}]{Mishra-Zhou-2004}
Mishra, B. and Zhou, Y. (2004).
\newblock \emph{Models of genome evolution} (Springer Verlag).
\newblock Natural Computing Series, Lecture Notes in Computer Science. 287--304
\bibAnnoteFile{Mishra-Zhou-2004}

\bibitem[{Moore et~al.(2006)Moore, Ghoshal, and Newman}]{MR2282142}
Moore, C., Ghoshal, G., and Newman, M. E.~J. (2006).
\newblock Exact solutions for models of evolving networks with addition and
  deletion of nodes.
\newblock \emph{Phys. Rev. E (3)} 74, 036121, 8.
\newblock \doi{10.1103/PhysRevE.74.036121}
\bibAnnoteFile{MR2282142}

\bibitem[{Newman(2003{\natexlab{a}})}]{homo1}
Newman, M. (2003{\natexlab{a}}).
\newblock Assortative mixing in networks.
\newblock \emph{Phys Rev Lett} 89, 758--762
\bibAnnoteFile{homo1}

\bibitem[{Newman et~al.(2011)Newman, Barabasi, and Watts}]{newman2011structure}
Newman, M., Barabasi, A.-L., and Watts, D.~J. (2011).
\newblock \emph{The structure and dynamics of networks} (Princeton University
  Press)
\bibAnnoteFile{newman2011structure}

\bibitem[{Newman(2001)}]{MR1812610}
Newman, M. E.~J. (2001).
\newblock The structure of scientific collaboration networks.
\newblock \emph{Proc. Natl. Acad. Sci. USA} 98, 404--409.
\newblock \doi{10.1073/pnas.021544898}
\bibAnnoteFile{MR1812610}

\bibitem[{Newman(2003{\natexlab{b}})}]{MR1975193}
Newman, M. E.~J. (2003{\natexlab{b}}).
\newblock Mixing patterns in networks.
\newblock \emph{Phys. Rev. E (3)} 67, 026126, 13.
\newblock \doi{10.1103/PhysRevE.67.026126}
\bibAnnoteFile{MR1975193}

\bibitem[{Newman(2003{\natexlab{c}})}]{MR2016118}
Newman, M. E.~J. (2003{\natexlab{c}}).
\newblock Random graphs as models of networks.
\newblock In \emph{Handbook of graphs and networks} (Wiley-VCH, Weinheim).
  35--68
\bibAnnoteFile{MR2016118}

\bibitem[{Newman(2003{\natexlab{d}})}]{MR2010377}
Newman, M. E.~J. (2003{\natexlab{d}}).
\newblock The structure and function of complex networks.
\newblock \emph{SIAM Rev.} 45, 167--256.
\newblock \doi{10.1137/S003614450342480}
\bibAnnoteFile{MR2010377}

\bibitem[{Newman(2004)}]{MR2108985}
Newman, M. E.~J. (2004).
\newblock Who is the best connected scientist? {A} study of scientific
  coauthorship networks.
\newblock In \emph{Complex networks} (Springer, Berlin), vol. 650 of
  \emph{Lecture Notes in Phys.} 337--370.
\newblock \doi{10.1007/978-3-540-44485-5_16}
\bibAnnoteFile{MR2108985}

\bibitem[{Newman(2006)}]{MR2282139}
Newman, M. E.~J. (2006).
\newblock Finding community structure in networks using the eigenvectors of
  matrices.
\newblock \emph{Phys. Rev. E (3)} 74, 036104, 19.
\newblock \doi{10.1103/PhysRevE.74.036104}
\bibAnnoteFile{MR2282139}

\bibitem[{Newman(2010)}]{MR2676073}
Newman, M. E.~J. (2010).
\newblock \emph{Networks} (Oxford University Press, Oxford).
\newblock \doi{10.1093/acprof:oso/9780199206650.001.0001}.
\newblock An introduction
\bibAnnoteFile{MR2676073}

\bibitem[{Newman and Watts(1999)}]{MR1732095}
Newman, M. E.~J. and Watts, D.~J. (1999).
\newblock Renormalization group analysis of the small-world network model.
\newblock \emph{Phys. Lett. A} 263, 341--346.
\newblock \doi{10.1016/S0375-9601(99)00757-4}
\bibAnnoteFile{MR1732095}

\bibitem[{Nogales et~al.(2017)Nogales, Louder, and He}]{tf}
Nogales, E., Louder, R., and He, Y. (2017).
\newblock Structural insights into the eukaryotic transcription initiation
  machinery.
\newblock \emph{Ann Rev Biophys} 46, 59--83
\bibAnnoteFile{tf}

\bibitem[{Noh(2003)}]{PhysRevE.67.045103}
Noh, J.~D. (2003).
\newblock Exact scaling properties of a hierarchical network model.
\newblock \emph{Phys. Rev. E} 67, 045103.
\newblock \doi{10.1103/PhysRevE.67.045103}
\bibAnnoteFile{PhysRevE.67.045103}

\bibitem[{Ohno(1970)}]{ohno}
Ohno, S. (1970).
\newblock \emph{Evolution by gene duplication} (Berlin: Springer-Verlag)
\bibAnnoteFile{ohno}

\bibitem[{Pellegrini(2019)}]{PELLEGRINI2019}
Pellegrini, M. (2019).
\newblock Community detection in biological networks.
\newblock In \emph{Encyclopedia of Bioinformatics and Computational Biology},
  eds. S.~Ranganathan, M.~Gribskov, K.~Nakai, and C.~Schönbach (Oxford:
  Academic Press). 978 -- 987.
\newblock \doi{https://doi.org/10.1016/B978-0-12-809633-8.20428-7}
\bibAnnoteFile{PELLEGRINI2019}

\bibitem[{Pinter et~al.(2005)Pinter, Rokhlenko, Yeger-Lotem, and
  Ziv-Ukelson}]{Pinter2005}
Pinter, R.~Y., Rokhlenko, O., Yeger-Lotem, E., and Ziv-Ukelson, M. (2005).
\newblock Alignment of metabolic pathways.
\newblock \emph{Bioinformatics} 21, 3401--3408.
\newblock \doi{10.1093/bioinformatics/bti554}
\bibAnnoteFile{Pinter2005}

\bibitem[{Piraveenan et~al.(2012)Piraveenan, Prokopenko, and
  Zomaya}]{MR2982456}
Piraveenan, M., Prokopenko, M., and Zomaya, A.~Y. (2012).
\newblock On congruity of nodes and assortative information content in complex
  networks.
\newblock \emph{Netw. Heterog. Media} 7, 441--461.
\newblock \doi{10.3934/nhm.2012.7.441}
\bibAnnoteFile{MR2982456}

\bibitem[{Poulos et~al.(2015)Poulos, Sloane, Hesson, and Wong}]{cis}
Poulos, R., Sloane, M., Hesson, L., and Wong, J. (2015).
\newblock The search for \textit{cis}-regulatory driver mutations in cancer
  genomes.
\newblock \emph{Oncotarget} 6, 32509--32525
\bibAnnoteFile{cis}

\bibitem[{Ravasz and Barab\'asi(2003)}]{ravasz03-1}
Ravasz, E. and Barab\'asi, A.-L. (2003).
\newblock Hierarchical organization in complex networks.
\newblock \emph{Physical Review E} 67, 026112
\bibAnnoteFile{ravasz03-1}

\bibitem[{Rodgers and McVey(2016)}]{imprecise}
Rodgers, K. and McVey, M. (2016).
\newblock Error-prone repair of dna double-strand breaks.
\newblock \emph{J Cell Physiol} 231, 15--24
\bibAnnoteFile{imprecise}

\bibitem[{Sarnak(2004)}]{Sarnak:AMS}
Sarnak, P. (2004).
\newblock What is{$\dots$}an expander?
\newblock \emph{Notices Amer. Math. Soc.} 51, 762--763
\bibAnnoteFile{Sarnak:AMS}

\bibitem[{Schuster et~al.(1994)Schuster, Fontana, Stadler, and
  Hofacker}]{neutrna}
Schuster, P., Fontana, W., Stadler, P., and Hofacker, I. (1994).
\newblock From sequences to shapes and back: A case-study in rna secondary
  structures.
\newblock \emph{Proc R Soc Lond B} 255, 279--284
\bibAnnoteFile{neutrna}

\bibitem[{Schwartz et~al.(2002)Schwartz, Cohen, ben Avraham, Barab\'{a}si, and
  Havlin}]{MR1919733}
Schwartz, N., Cohen, R., ben Avraham, D., Barab\'{a}si, A.-L., and Havlin, S.
  (2002).
\newblock Percolation in directed scale-free networks.
\newblock \emph{Phys. Rev. E (3)} 66, 015104, 4.
\newblock \doi{10.1103/PhysRevE.66.015104}
\bibAnnoteFile{MR1919733}

\bibitem[{Serre(1980)}]{serre2}
Serre, J.-P. (1980).
\newblock \emph{Trees} (Berlin: Springer-Verlag).
\newblock Translated from the French by John Stillwell
\bibAnnoteFile{serre2}

\bibitem[{Shapley(1969)}]{shap}
Shapley, L. (1969).
\newblock A value for n person games.
\newblock In \emph{The Shapley Value} (Cambridge: Cambridge University Press)
\bibAnnoteFile{shap}

\bibitem[{Sharan et~al.(2005)Sharan, Suthram, Kelley, Kuhn, McCuine, Uetz
  et~al.}]{Sharan1974}
Sharan, R., Suthram, S., Kelley, R.~M., Kuhn, T., McCuine, S., Uetz, P., et~al.
  (2005).
\newblock Conserved patterns of protein interaction in multiple species.
\newblock \emph{Proceedings of the National Academy of Sciences} 102,
  1974--1979.
\newblock \doi{10.1073/pnas.0409522102}
\bibAnnoteFile{Sharan1974}

\bibitem[{Smeenk et~al.(2008)Smeenk, van Heeringen, Koeppel, van Driel,
  Bartels, Akkers et~al.}]{binding}
Smeenk, L., van Heeringen, S., Koeppel, M., van Driel, M., Bartels, S., Akkers,
  R., et~al. (2008).
\newblock Characterization of genome-wide p53-binding sites upon stress
  binding.
\newblock \emph{Nuc Acids Res} 36, 3639--3654
\bibAnnoteFile{binding}

\bibitem[{Song et~al.(2010)Song, Qu, Blumm, and Barab\'{a}si}]{MR2643139}
Song, C., Qu, Z., Blumm, N., and Barab\'{a}si, A.-L. (2010).
\newblock Limits of predictability in human mobility.
\newblock \emph{Science} 327, 1018--1021.
\newblock \doi{10.1126/science.1177170}
\bibAnnoteFile{MR2643139}

\bibitem[{Spielman(2018)}]{Spielman-LectNotes}
Spielman, D. (2018).
\newblock \emph{Spectral Graph Theory and Its Applications}
  (http://www.cs.yale.edu/homes/spielman)
\bibAnnoteFile{Spielman-LectNotes}

\bibitem[{Spielman(1996)}]{MR1465732}
Spielman, D.~A. (1996).
\newblock Linear-time encodable and decodable error-correcting codes.
\newblock \emph{IEEE Trans. Inform. Theory} 42, 1723--1731.
\newblock \doi{10.1109/18.556668}.
\newblock Codes and complexity
\bibAnnoteFile{MR1465732}

\bibitem[{Spielman and Teng(2004)}]{Spielman-Teng-2004}
Spielman, D.~A. and Teng, S.-H. (2004).
\newblock Smoothed analysis of algorithms: why the simplex algorithm usually
  takes polynomial time.
\newblock \emph{J. ACM} 51, 385--463.
\newblock \doi{10.1145/990308.990310}
\bibAnnoteFile{Spielman-Teng-2004}

\bibitem[{Spielman and Teng(2011{\natexlab{a}})}]{Spielman-Teng-2011}
Spielman, D.~A. and Teng, S.-H. (2011{\natexlab{a}}).
\newblock Spectral sparsification of graphs.
\newblock \emph{SIAM J. Comput.} 40, 981--1025.
\newblock \doi{10.1137/08074489X}
\bibAnnoteFile{Spielman-Teng-2011}

\bibitem[{Spielman and Teng(2011{\natexlab{b}})}]{MR2825307}
Spielman, D.~A. and Teng, S.-H. (2011{\natexlab{b}}).
\newblock Spectral sparsification of graphs.
\newblock \emph{SIAM J. Comput.} 40, 981--1025.
\newblock \doi{10.1137/08074489X}
\bibAnnoteFile{MR2825307}

\bibitem[{Spielman and Teng(2013)}]{Spielman-Teng}
Spielman, D.~A. and Teng, S.-H. (2013).
\newblock A local clustering algorithm for massive graphs and its application
  to nearly linear time graph partitioning.
\newblock \emph{SIAM J. Comput.} 42, 1--26.
\newblock \doi{10.1137/080744888}
\bibAnnoteFile{Spielman-Teng}

\bibitem[{Spielman and Teng(2014)}]{Spielman-Teng2014}
Spielman, D.~A. and Teng, S.-H. (2014).
\newblock Nearly linear time algorithms for preconditioning and solving
  symmetric, diagonally dominant linear systems.
\newblock \emph{SIAM J. Matrix Anal. Appl.} 35, 835--885.
\newblock \doi{10.1137/090771430}
\bibAnnoteFile{Spielman-Teng2014}

\bibitem[{Taylor and Jonker(1978)}]{tj1}
Taylor, P. and Jonker, L. (1978).
\newblock Evolutionary stable strategies and game dynamics.
\newblock \emph{Mathematical Biosciences} 40, 145--156
\bibAnnoteFile{tj1}

\bibitem[{Thompson et~al.(2015)Thompson, Regev, and Roy}]{genereg}
Thompson, D., Regev, A., and Roy, S. (2015).
\newblock Comparative analysis of gene regulatory networks: from network
  reconstruction to evolution.
\newblock \emph{Annual Review of Cell and Developmental Biology} 31, 399--428
\bibAnnoteFile{genereg}

\bibitem[{Thulasiraman et~al.(2015)Thulasiraman, Arumugam, Brandstadt, and
  Nishizeki}]{thulasiraman2015}
Thulasiraman, K., Arumugam, S., Brandstadt, A., and Nishizeki, T. (2015).
\newblock \emph{Handbook of Graph Theory, Combinatorial Optimization, and
  Algorithms}.
\newblock Chapman \& Hall/CRC Computer and Information Science Series (Taylor
  \& Francis)
\bibAnnoteFile{thulasiraman2015}

\bibitem[{Vazquez et~al.(2008)Vazquez, de~Menezes, Barab\'{a}si, and
  Oltvai}]{MR2457138}
Vazquez, A., de~Menezes, M.~A., Barab\'{a}si, A.-L., and Oltvai, Z.~N. (2008).
\newblock Impact of limited solvent capacity on metabolic rate, enzyme
  activities, and metabolite concentrations of {\it {s}. cerevisiae}
  glycolysis.
\newblock \emph{PLoS Comput. Biol.} 4, e1000195, 6.
\newblock \doi{10.1371/journal.pcbi.1000195}
\bibAnnoteFile{MR2457138}

\bibitem[{Wagner(1994)}]{wag}
Wagner, A. (1994).
\newblock Evolution of gene networks by gene duplications: a mathematical model
  and its implications on genome organization.
\newblock \emph{Proc Natl Acad Sci USA} 91, 4387--4391
\bibAnnoteFile{wag}

\bibitem[{Watts(1999)}]{MR1716136}
Watts, D.~J. (1999).
\newblock \emph{Small worlds}.
\newblock Princeton Studies in Complexity (Princeton University Press,
  Princeton, NJ).
\newblock The dynamics of networks between order and randomness
\bibAnnoteFile{MR1716136}

\bibitem[{Watts(2003)}]{MR2041642}
Watts, D.~J. (2003).
\newblock \emph{Six degrees} (W. W. Norton \& Co. Inc., New York).
\newblock The science of a connected age
\bibAnnoteFile{MR2041642}

\bibitem[{Watts and Strogatz(1998)}]{Watts1998CollectiveDO}
Watts, D.~J. and Strogatz, S.~H. (1998).
\newblock Collective dynamics of ‘small-world’ networks.
\newblock \emph{Nature} 393, 440--442
\bibAnnoteFile{Watts1998CollectiveDO}

\bibitem[{Yamada et~al.(2011)Yamada, Letunic, Okuda, Kanehisa, and Bork}]{meta}
Yamada, T., Letunic, I., Okuda, S., Kanehisa, M., and Bork, P. (2011).
\newblock ipath2.0: interactive pathway explorer.
\newblock \emph{Nuc Acids Res} 39, W412--W415
\bibAnnoteFile{meta}

\bibitem[{Zhang et~al.(2016)Zhang, Moore, and Newman}]{MR3678034}
Zhang, P., Moore, C., and Newman, M. E.~J. (2016).
\newblock Community detection in networks with unequal groups.
\newblock \emph{Phys. Rev. E} 93, 012303, 12.
\newblock \doi{10.1103/physreve.93.012303}
\bibAnnoteFile{MR3678034}

\bibitem[{Zhang et~al.(2017)Zhang, Moore, and Newman}]{MR3713556}
Zhang, X., Moore, C., and Newman, M. E.~J. (2017).
\newblock Random graph models for dynamic networks.
\newblock \emph{Eur. Phys. J. B} 90, Paper No. 200, 14.
\newblock \doi{10.1140/epjb/e2017-80122-8}
\bibAnnoteFile{MR3713556}

\bibitem[{Zhang et~al.(2005)Zhang, Luo, Kishino, and Kearsey}]{power}
Zhang, Z., Luo, Z., Kishino, H., and Kearsey, M. (2005).
\newblock Divergence pattern of duplicate genes in protein-protein interactions
  follows the power law.
\newblock \emph{Mol Biol Evol} 22, 501--505
\bibAnnoteFile{power}

\bibitem[{Zhou(2005)}]{Zhou05}
Zhou, Y.~J. (2005).
\newblock \emph{Statistical Analyses and Markov Modeling of Duplication in
  Genome Evolution}.
\newblock Thesis (Ph.D.)--New York University (NYU)
\bibAnnoteFile{Zhou05}

\end{thebibliography}


\begin{thebibliography}{73}
\providecommand{\natexlab}[1]{#1}
\expandafter\ifx\csname urlstyle\endcsname\relax
  \providecommand{\doi}[1]{doi:\discretionary{}{}{}#1}\else
  \providecommand{\doi}{doi:\discretionary{}{}{}\begingroup
  \urlstyle{rm}\Url}\fi
\providecommand{\selectlanguage}[1]{\relax}
\providecommand{\bibAnnoteFile}[1]{%
  \IfFileExists{#1}{\begin{quotation}\noindent\textsc{Key:} #1\\
  \textsc{Annotation:}\ \input{#1}\end{quotation}}{}}
\providecommand{\bibAnnote}[2]{%
  \begin{quotation}\noindent\textsc{Key:} #1\\
  \textsc{Annotation:}\ #2\end{quotation}}

\bibitem[{Alderson(2008)}]{MR2468898}
Alderson, D.~L. (2008).
\newblock Catching the ``network science'' bug: insight and opportunity for the
  operations researcher.
\newblock \emph{Oper. Res.} 56, 1047--1065.
\newblock \doi{10.1287/opre.1080.0606}
\bibAnnoteFile{MR2468898}

\bibitem[{Alon and Spencer(1992)}]{MR1140703}
Alon, N. and Spencer, J.~H. (1992).
\newblock \emph{The probabilistic method}.
\newblock Wiley-Interscience Series in Discrete Mathematics and Optimization
  (New York: John Wiley \& Sons Inc.).
\newblock With an appendix by Paul Erd{\H{o}}s, A Wiley-Interscience
  Publication
\bibAnnoteFile{MR1140703}

\bibitem[{Andersen et~al.(2008)Andersen, Chung, and Lang}]{MR2560260}
Andersen, R., Chung, F., and Lang, K. (2008).
\newblock Local partitioning for directed graphs using {P}age{R}ank.
\newblock \emph{Internet Math.} 5, 3--22
\bibAnnoteFile{MR2560260}

\bibitem[{Andersen et~al.(2007)Andersen, Chung, and Lu}]{MR2304986}
Andersen, R., Chung, F., and Lu, L. (2007).
\newblock Drawing power law graphs using a local/global decomposition.
\newblock \emph{Algorithmica} 47, 379--397.
\newblock \doi{10.1007/s00453-006-0160-2}
\bibAnnoteFile{MR2304986}

\bibitem[{Babai(1979)}]{MR546860}
Babai, L. (1979).
\newblock Spectra of {C}ayley graphs.
\newblock \emph{J. Combin. Theory Ser. B} 27, 180--189.
\newblock \doi{10.1016/0095-8956(79)90079-0}
\bibAnnoteFile{MR546860}

\bibitem[{Banerjee and Jost(2009)}]{MR2527959}
Banerjee, A. and Jost, J. (2009).
\newblock Graph spectra as a systematic tool in computational biology.
\newblock \emph{Discrete Appl. Math.} 157, 2425--2431.
\newblock \doi{10.1016/j.dam.2008.06.033}
\bibAnnoteFile{MR2527959}

\bibitem[{Barg and Z\'{e}mor(2004)}]{Barg:Zemor}
Barg, A. and Z\'{e}mor, G. (2004).
\newblock Error exponents of expander codes under linear-complexity decoding.
\newblock \emph{SIAM J. Discrete Math.} 17, 426--445.
\newblock \doi{10.1137/S0895480102403799}
\bibAnnoteFile{Barg:Zemor}

\bibitem[{Beyer(2009)}]{MR2759501}
Beyer, A. (2009).
\newblock Network-based models in molecular biology.
\newblock In \emph{Dynamics on and of complex networks} (Birkh\"{a}user Boston,
  Inc., Boston, MA), Model. Simul. Sci. Eng. Technol. 35--56.
\newblock \doi{10.1007/978-0-8176-4751-3_3}
\bibAnnoteFile{MR2759501}

\bibitem[{Bien(1989)}]{MR972207}
Bien, F. (1989).
\newblock Constructions of telephone networks by group representations.
\newblock \emph{Notices Amer. Math. Soc.} 36, 5--22
\bibAnnoteFile{MR972207}

\bibitem[{Biggs(1993)}]{biggs}
Biggs, N. (1993).
\newblock \emph{Algebraic graph theory}.
\newblock Cambridge Mathematical Library (Cambridge: Cambridge University
  Press), second edn.
\bibAnnoteFile{biggs}

\bibitem[{Bollob\'{a}s(1978)}]{bollobas1}
Bollob\'{a}s, B. (1978).
\newblock \emph{Extremal graph theory}, vol.~11 of \emph{London Mathematical
  Society Monographs} (Academic Press, Inc. [Harcourt Brace Jovanovich,
  Publishers], London-New York)
\bibAnnoteFile{bollobas1}

\bibitem[{Bollob{\'a}s(1978)}]{MR506522}
Bollob{\'a}s, B. (1978).
\newblock \emph{Extremal graph theory}, vol.~11 of \emph{London Mathematical
  Society Monographs} (London: Academic Press Inc. [Harcourt Brace Jovanovich
  Publishers])
\bibAnnoteFile{MR506522}

\bibitem[{Bollob{\'a}s(1986)}]{MR840466}
Bollob{\'a}s, B. (1986).
\newblock \emph{Extremal graph theory with emphasis on probabilistic methods},
  vol.~62 of \emph{CBMS Regional Conference Series in Mathematics} (Published
  for the Conference Board of the Mathematical Sciences, Washington, DC)
\bibAnnoteFile{MR840466}

\bibitem[{Bollob{\'a}s(1991)}]{MR1141921}
Bollob{\'a}s, B. (1991).
\newblock Random graphs.
\newblock In \emph{Probabilistic combinatorics and its applications ({S}an
  {F}rancisco, {CA}, 1991)} (Providence, RI: Amer. Math. Soc.), vol.~44 of
  \emph{Proc. Sympos. Appl. Math.} 1--20
\bibAnnoteFile{MR1141921}

\bibitem[{Brouwer et~al.(1989)Brouwer, Cohen, and Neumaier}]{MR1002568}
Brouwer, A.~E., Cohen, A.~M., and Neumaier, A. (1989).
\newblock \emph{Distance-regular graphs}, vol.~18 of \emph{Ergebnisse der
  Mathematik und ihrer Grenzgebiete (3) [Results in Mathematics and Related
  Areas (3)]} (Berlin: Springer-Verlag)
\bibAnnoteFile{MR1002568}

\bibitem[{Chatterjee et~al.(2011)Chatterjee, Diaconis, and Sly}]{MR2857452}
Chatterjee, S., Diaconis, P., and Sly, A. (2011).
\newblock Random graphs with a given degree sequence.
\newblock \emph{Ann. Appl. Probab.} 21, 1400--1435.
\newblock \doi{10.1214/10-AAP728}
\bibAnnoteFile{MR2857452}

\bibitem[{Chung(2009)}]{MR2798107}
Chung, F. (2009).
\newblock A local graph partitioning algorithm using heat kernel {P}age{R}ank.
\newblock \emph{Internet Math.} 6, 315--330 (2010).
\newblock \doi{10.1080/15427951.2009.10390643}
\bibAnnoteFile{MR2798107}

\bibitem[{Chung and Lu(2004)}]{MR2108973}
Chung, F. and Lu, L. (2004).
\newblock The small world phenomenon in hybrid power law graphs.
\newblock In \emph{Complex networks} (Springer, Berlin), vol. 650 of
  \emph{Lecture Notes in Phys.} 89--104.
\newblock \doi{10.1007/978-3-540-44485-5_4}
\bibAnnoteFile{MR2108973}

\bibitem[{Chung and Lu(2006)}]{MR2248695}
Chung, F. and Lu, L. (2006).
\newblock \emph{Complex graphs and networks}, vol. 107 of \emph{CBMS Regional
  Conference Series in Mathematics} (Published for the Conference Board of the
  Mathematical Sciences, Washington, DC; by the American Mathematical Society,
  Providence, RI).
\newblock \doi{10.1090/cbms/107}
\bibAnnoteFile{MR2248695}

\bibitem[{Chung(1966)}]{Chung-Shape}
Chung, F. R.~K. (1966).
\newblock Can you hear the shape of a graph?
\newblock \emph{Amer. Math. Monthly}
\bibAnnoteFile{Chung-Shape}

\bibitem[{Chung(1991)}]{MR1141922}
Chung, F. R.~K. (1991).
\newblock Constructing random-like graphs.
\newblock In \emph{Probabilistic combinatorics and its applications ({S}an
  {F}rancisco, {CA}, 1991)} (Providence, RI: Amer. Math. Soc.), vol.~44 of
  \emph{Proc. Sympos. Appl. Math.} 21--55
\bibAnnoteFile{MR1141922}

\bibitem[{Chung(1997)}]{MR1421568}
Chung, F. R.~K. (1997).
\newblock \emph{Spectral graph theory}, vol.~92 of \emph{CBMS Regional
  Conference Series in Mathematics} (Published for the Conference Board of the
  Mathematical Sciences, Washington, DC; by the American Mathematical Society,
  Providence, RI)
\bibAnnoteFile{MR1421568}

\bibitem[{Chung et~al.(1988)Chung, Graham, and Wilson}]{MR928566}
Chung, F. R.~K., Graham, R.~L., and Wilson, R.~M. (1988).
\newblock Quasirandom graphs.
\newblock \emph{Proc. Nat. Acad. Sci. U.S.A.} 85, 969--970.
\newblock \doi{10.1073/pnas.85.4.969}
\bibAnnoteFile{MR928566}

\bibitem[{Cvetkovi{\'c} et~al.(1988)Cvetkovi{\'c}, Doob, Gutman, and
  Torga{\v{s}}ev}]{MR926481}
Cvetkovi{\'c}, D.~M., Doob, M., Gutman, I., and Torga{\v{s}}ev, A. (1988).
\newblock \emph{Recent results in the theory of graph spectra}, vol.~36 of
  \emph{Annals of Discrete Mathematics} (Amsterdam: North-Holland Publishing
  Co.)
\bibAnnoteFile{MR926481}

\bibitem[{Cvetkovi{\'c} et~al.(1980)Cvetkovi{\'c}, Doob, and Sachs}]{MR572262}
Cvetkovi{\'c}, D.~M., Doob, M., and Sachs, H. (1980).
\newblock \emph{Spectra of graphs}, vol.~87 of \emph{Pure and Applied
  Mathematics} (New York: Academic Press Inc. [Harcourt Brace Jovanovich
  Publishers]).
\newblock Theory and application
\bibAnnoteFile{MR572262}

\bibitem[{Davidoff et~al.(2003)Davidoff, Sarnak, and Valette}]{MR1989434}
Davidoff, G., Sarnak, P., and Valette, A. (2003).
\newblock \emph{Elementary number theory, group theory, and {R}amanujan
  graphs}, vol.~55 of \emph{London Mathematical Society Student Texts}
  (Cambridge University Press, Cambridge).
\newblock \doi{10.1017/CBO9780511615825}
\bibAnnoteFile{MR1989434}

\bibitem[{Delgado and Janwa(2017)}]{MR3600878}
Delgado, M. and Janwa, H. (2017).
\newblock On the conjecture on {APN} functions and absolute irreducibility of
  polynomials.
\newblock \emph{Des. Codes Cryptogr.} 82, 617--627.
\newblock \doi{10.1007/s10623-015-0168-1}
\bibAnnoteFile{MR3600878}

\bibitem[{Diaconis(1988)}]{MR964069}
Diaconis, P. (1988).
\newblock \emph{Group representations in probability and statistics}.
\newblock Institute of Mathematical Statistics Lecture Notes---Monograph
  Series, 11 (Hayward, CA: Institute of Mathematical Statistics)
\bibAnnoteFile{MR964069}

\bibitem[{Diaconis(1991)}]{MR1141927}
Diaconis, P. (1991).
\newblock Finite {F}ourier methods: access to tools.
\newblock In \emph{Probabilistic combinatorics and its applications ({S}an
  {F}rancisco, {CA}, 1991)} (Providence, RI: Amer. Math. Soc.), vol.~44 of
  \emph{Proc. Sympos. Appl. Math.} 171--194
\bibAnnoteFile{MR1141927}

\bibitem[{Dougherty and Janwa(1991)}]{MR1079013}
Dougherty, R. and Janwa, H. (1991).
\newblock Covering radius computations for binary cyclic codes.
\newblock \emph{Math. Comp.} 57, 415--434.
\newblock \doi{10.2307/2938683}.
\newblock With microfiche supplement
\bibAnnoteFile{MR1079013}

\bibitem[{Easley and Kleinberg(2010)}]{MR2677125}
Easley, D. and Kleinberg, J. (2010).
\newblock \emph{Networks, crowds, and markets} (Cambridge University Press,
  Cambridge).
\newblock \doi{10.1017/CBO9780511761942}.
\newblock Reasoning about a highly connected world
\bibAnnoteFile{MR2677125}

\bibitem[{Ganguly et~al.(2009)Ganguly, Deutsch, and Mukherjee}]{MR2759498}
Ganguly, N., Deutsch, A., and Mukherjee, A. (eds.) (2009).
\newblock \emph{Dynamics on and of complex networks}.
\newblock Modeling and Simulation in Science, Engineering and Technology
  (Birkh\"{a}user Boston, Inc., Boston, MA).
\newblock \doi{10.1007/978-0-8176-4751-3}.
\newblock Applications to biology, computer science, and the social sciences
\bibAnnoteFile{MR2759498}

\bibitem[{Gleich(2015)}]{MR3376760}
Gleich, D.~F. (2015).
\newblock Page{R}ank beyond the web.
\newblock \emph{SIAM Rev.} 57, 321--363.
\newblock \doi{10.1137/140976649}
\bibAnnoteFile{MR3376760}

\bibitem[{Guruswami and Indyk(2001)}]{Guruswamy:Indyk}
Guruswami, V. and Indyk, P. (2001).
\newblock Expander-based constructions of efficiently decodable codes (extended
  abstract).
\newblock In \emph{42nd {IEEE} {S}ymposium on {F}oundations of {C}omputer
  {S}cience ({L}as {V}egas, {NV}, 2001)} (IEEE Computer Soc., Los Alamitos,
  CA). 658--667
\bibAnnoteFile{Guruswamy:Indyk}

\bibitem[{H\o~holdt and Janwa(2012)}]{MR2988185}
H\o~holdt, T. and Janwa, H. (2012).
\newblock Eigenvalues and expansion of bipartite graphs.
\newblock \emph{Des. Codes Cryptogr.} 65, 259--273.
\newblock \doi{10.1007/s10623-011-9598-6}
\bibAnnoteFile{MR2988185}

\bibitem[{H\o~holdt and Janwal(2009)}]{MR2580853}
H\o~holdt, T. and Janwal, H. (2009).
\newblock Optimal bipartite {R}amanujan graphs from balanced incomplete block
  designs: their characterizations and applications to expander/{LDPC} codes.
\newblock In \emph{Applied algebra, algebraic algorithms, and error-correcting
  codes} (Springer, Berlin), vol. 5527 of \emph{Lecture Notes in Comput. Sci.}
  53--64.
\newblock \doi{10.1007/978-3-642-02181-7_6}
\bibAnnoteFile{MR2580853}

\bibitem[{Horn and Johnson(2013)}]{MR2978290}
Horn, R.~A. and Johnson, C.~R. (2013).
\newblock \emph{Matrix analysis} (Cambridge University Press, Cambridge),
  second edn.
\bibAnnoteFile{MR2978290}

\bibitem[{Janwa(2003)}]{MR2042419}
Janwa, H. (2003).
\newblock Good expander graphs and expander codes: parameters and decoding.
\newblock In \emph{Applied algebra, algebraic algorithms and error-correcting
  codes ({T}oulouse, 2003)} (Springer, Berlin), vol. 2643 of \emph{Lecture
  Notes in Comput. Sci.} 119--128.
\newblock \doi{10.1007/3-540-44828-4_14}
\bibAnnoteFile{MR2042419}

\bibitem[{Janwa and Lal(2003)}]{MR1959169}
Janwa, H. and Lal, A.~K. (2003).
\newblock On {T}anner codes: minimum distance and decoding.
\newblock \emph{Appl. Algebra Engrg. Comm. Comput.} 13, 335--347.
\newblock \doi{10.1007/s00200-003-0098-4}
\bibAnnoteFile{MR1959169}

\bibitem[{Janwa et~al.(1995)Janwa, McGuire, and Wilson}]{MR1359909}
Janwa, H., McGuire, G.~M., and Wilson, R.~M. (1995).
\newblock Double-error-correcting cyclic codes and absolutely irreducible
  polynomials over {${\rm GF}(2)$}.
\newblock \emph{J. Algebra} 178, 665--676.
\newblock \doi{10.1006/jabr.1995.1372}
\bibAnnoteFile{MR1359909}

\bibitem[{Janwa and Moreno(1996)}]{MR1403868}
Janwa, H. and Moreno, O. (1996).
\newblock Mc{E}liece public key cryptosystems using algebraic-geometric codes.
\newblock \emph{Des. Codes Cryptogr.} 8, 293--307.
\newblock \doi{10.1023/A:1027351723034}
\bibAnnoteFile{MR1403868}

\bibitem[{Janwa and Moreno(1998)}]{MR1676451}
Janwa, H. and Moreno, O. (1998).
\newblock Coding-theoretic constructions of some number-theoretic {R}amanujan
  graphs.
\newblock In \emph{Proceedings of the {T}wenty-ninth {S}outheastern
  {I}nternational {C}onference on {C}ombinatorics, {G}raph {T}heory and
  {C}omputing ({B}oca {R}aton, {FL}, 1998)}. vol. 130, 63--76
\bibAnnoteFile{MR1676451}

\bibitem[{Janwa and Rangachari(2015)}]{MR37}
Janwa, H. and Rangachari, S. (2015).
\newblock Ramanujan graphs and their applications
\bibAnnoteFile{MR37}

\bibitem[{Kac(1966)}]{MR0201237}
Kac, M. (1966).
\newblock Can one hear the shape of a drum?
\newblock \emph{Amer. Math. Monthly} 73, 1--23.
\newblock \doi{10.2307/2313748}
\bibAnnoteFile{MR0201237}

\bibitem[{Levin and Peres(2017)}]{MR3726904}
Levin, D.~A. and Peres, Y. (2017).
\newblock \emph{Markov chains and mixing times} (American Mathematical Society,
  Providence, RI).
\newblock Second edition of [ MR2466937], With contributions by Elizabeth L.
  Wilmer, With a chapter on ``Coupling from the past'' by James G. Propp and
  David B. Wilson
\bibAnnoteFile{MR3726904}

\bibitem[{Li(1996)}]{MR1390759}
Li, W. C.~W. (1996).
\newblock \emph{Number theory with applications}, vol.~7 of \emph{Series on
  University Mathematics} (River Edge, NJ: World Scientific Publishing Co.
  Inc.)
\bibAnnoteFile{MR1390759}

\bibitem[{Lov\'{a}sz et~al.(1999)Lov\'{a}sz, Gy\'{a}rf\'{a}s, Katona, Recski,
  and Sz\'{e}kely}]{MR1673501}
Lov\'{a}sz, L., Gy\'{a}rf\'{a}s, A., Katona, G., Recski, A., and Sz\'{e}kely,
  L. (eds.) (1999).
\newblock \emph{Graph theory and combinatorial biology}, vol.~7 of \emph{Bolyai
  Society Mathematical Studies} (J\'{a}nos Bolyai Mathematical Society,
  Budapest)
\bibAnnoteFile{MR1673501}

\bibitem[{Lubotzky(1994)}]{MR1308046}
Lubotzky, A. (1994).
\newblock \emph{Discrete groups, expanding graphs and invariant measures}, vol.
  125 of \emph{Progress in Mathematics} (Basel: Birkh\"auser Verlag).
\newblock With an appendix by Jonathan D. Rogawski
\bibAnnoteFile{MR1308046}

\bibitem[{Lubotzky(2012)}]{MR2869010}
Lubotzky, A. (2012).
\newblock Expander graphs in pure and applied mathematics.
\newblock \emph{Bull. Amer. Math. Soc. (N.S.)} 49, 113--162.
\newblock \doi{10.1090/S0273-0979-2011-01359-3}
\bibAnnoteFile{MR2869010}

\bibitem[{Lubotzky et~al.(1988)Lubotzky, Phillips, and Sarnak}]{MR963118}
Lubotzky, A., Phillips, R., and Sarnak, P. (1988).
\newblock Ramanujan graphs.
\newblock \emph{Combinatorica} 8, 261--277.
\newblock \doi{10.1007/BF02126799}
\bibAnnoteFile{MR963118}

\bibitem[{MacWilliams and Sloane(1977)}]{MR0465510}
MacWilliams, F.~J. and Sloane, N. J.~A. (1977).
\newblock \emph{The theory of error-correcting codes. {II}} (Amsterdam:
  North-Holland Publishing Co.).
\newblock North-Holland Mathematical Library, Vol. 16
\bibAnnoteFile{MR0465510}

\bibitem[{Margulis(1988)}]{MR939574}
Margulis, G.~A. (1988).
\newblock Explicit group-theoretic constructions of combinatorial schemes and
  their applications in the construction of expanders and concentrators.
\newblock \emph{Problemy Peredachi Informatsii} 24, 51--60
\bibAnnoteFile{MR939574}

\bibitem[{Masuda et~al.(2017)Masuda, Porter, and Lambiotte}]{MR3730470}
Masuda, N., Porter, M.~A., and Lambiotte, R. (2017).
\newblock Random walks and diffusion on networks.
\newblock \emph{Phys. Rep.} 716/717, 1--58.
\newblock \doi{10.1016/j.physrep.2017.07.007}
\bibAnnoteFile{MR3730470}

\bibitem[{Paul et~al.(1976/77)Paul, Tarjan, and Celoni}]{MR451347}
Paul, W.~J., Tarjan, R.~E., and Celoni, J.~R. (1976/77).
\newblock Space bounds for a game on graphs.
\newblock \emph{Math. Systems Theory} 10, 239--251.
\newblock \doi{10.1007/BF01683275}
\bibAnnoteFile{MR451347}

\bibitem[{Pi\~nero and Janwa(2014)}]{MR3147592}
Pi\~nero, F. and Janwa, H. (2014).
\newblock On the subfield subcodes of {H}ermitian codes.
\newblock \emph{Des. Codes Cryptogr.} 70, 157--173.
\newblock \doi{10.1007/s10623-012-9736-9}
\bibAnnoteFile{MR3147592}

\bibitem[{Pippenger(1990{\natexlab{a}})}]{MR1127181}
Pippenger, N. (1990{\natexlab{a}}).
\newblock Communication networks.
\newblock In \emph{Handbook of theoretical computer science, {V}ol.\ {A}}
  (Amsterdam: Elsevier). 805--833
\bibAnnoteFile{MR1127181}

\bibitem[{Pippenger(1990{\natexlab{b}})}]{MR1081951}
Pippenger, N. (1990{\natexlab{b}}).
\newblock Selection networks.
\newblock In \emph{Algorithms ({T}okyo, 1990)} (Berlin: Springer), vol. 450 of
  \emph{Lecture Notes in Comput. Sci.} 2--11.
\newblock \doi{10.1007/3-540-52921-7_50}
\bibAnnoteFile{MR1081951}

\bibitem[{Pippenger and Lin(1994)}]{MR1259014}
Pippenger, N. and Lin, G. (1994).
\newblock Fault-tolerant circuit-switching networks.
\newblock \emph{SIAM J. Discrete Math.} 7, 108--118.
\newblock \doi{10.1137/S0895480192229790}
\bibAnnoteFile{MR1259014}

\bibitem[{Piraveenan et~al.(2012)Piraveenan, Prokopenko, and
  Zomaya}]{MR2982456}
Piraveenan, M., Prokopenko, M., and Zomaya, A.~Y. (2012).
\newblock On congruity of nodes and assortative information content in complex
  networks.
\newblock \emph{Netw. Heterog. Media} 7, 441--461.
\newblock \doi{10.3934/nhm.2012.7.441}
\bibAnnoteFile{MR2982456}

\bibitem[{Randles et~al.(2011)Randles, Lamb, Odat, and
  Taleb-Bendiab}]{MR2780129}
Randles, M., Lamb, D., Odat, E., and Taleb-Bendiab, A. (2011).
\newblock Distributed redundancy and robustness in complex systems.
\newblock \emph{J. Comput. System Sci.} 77, 293--304.
\newblock \doi{10.1016/j.jcss.2010.01.008}
\bibAnnoteFile{MR2780129}

\bibitem[{Sarnak(1990)}]{MR1102679}
Sarnak, P. (1990).
\newblock \emph{Some applications of modular forms}, vol.~99 of \emph{Cambridge
  Tracts in Mathematics} (Cambridge: Cambridge University Press).
\newblock \doi{10.1017/CBO9780511895593}
\bibAnnoteFile{MR1102679}

\bibitem[{Sarnak(2004)}]{Sarnak:AMS}
Sarnak, P. (2004).
\newblock What is{$\dots$}an expander?
\newblock \emph{Notices Amer. Math. Soc.} 51, 762--763
\bibAnnoteFile{Sarnak:AMS}

\bibitem[{Serre(1980)}]{serre2}
Serre, J.-P. (1980).
\newblock \emph{Trees} (Berlin: Springer-Verlag).
\newblock Translated from the French by John Stillwell
\bibAnnoteFile{serre2}

\bibitem[{Sipser and Spielman(1996)}]{MR1465731}
Sipser, M. and Spielman, D.~A. (1996).
\newblock Expander codes.
\newblock \emph{IEEE Trans. Inform. Theory} 42, 1710--1722.
\newblock \doi{10.1109/18.556667}.
\newblock Codes and complexity
\bibAnnoteFile{MR1465731}

\bibitem[{Sol\'{e} and Valverde(2004)}]{MR2108978}
Sol\'{e}, R.~V. and Valverde, S. (2004).
\newblock Information theory of complex networks: on evolution and
  architectural constraints.
\newblock In \emph{Complex networks} (Springer, Berlin), vol. 650 of
  \emph{Lecture Notes in Phys.} 189--207.
\newblock \doi{10.1007/978-3-540-44485-5_9}
\bibAnnoteFile{MR2108978}

\bibitem[{Spielman(1996)}]{MR1465732}
Spielman, D.~A. (1996).
\newblock Linear-time encodable and decodable error-correcting codes.
\newblock \emph{IEEE Trans. Inform. Theory} 42, 1723--1731.
\newblock \doi{10.1109/18.556668}.
\newblock Codes and complexity
\bibAnnoteFile{MR1465732}

\bibitem[{Takemoto and Oosawa(2007)}]{MR2348125}
Takemoto, K. and Oosawa, C. (2007).
\newblock Modeling for evolving biological networks with scale-free
  connectivity, hierarchical modularity, and disassortativity.
\newblock \emph{Math. Biosci.} 208, 454--468.
\newblock \doi{10.1016/j.mbs.2006.11.002}
\bibAnnoteFile{MR2348125}

\bibitem[{Tanner(1984)}]{MR752035}
Tanner, R.~M. (1984).
\newblock Explicit concentrators from generalized {$N$}-gons.
\newblock \emph{SIAM J. Algebraic Discrete Methods} 5, 287--293.
\newblock \doi{10.1137/0605030}
\bibAnnoteFile{MR752035}

\bibitem[{Tanner(2001)}]{Tanner:IEEE:2001}
Tanner, R.~M. (2001).
\newblock Minimum-distance bounds by graph analysis.
\newblock \emph{IEEE Trans. Inform. Theory} 47, 808--821.
\newblock \doi{10.1109/18.910591}
\bibAnnoteFile{Tanner:IEEE:2001}

\bibitem[{Terras(1999)}]{MR1695775}
Terras, A. (1999).
\newblock \emph{Fourier analysis on finite groups and applications}, vol.~43 of
  \emph{London Mathematical Society Student Texts} (Cambridge: Cambridge
  University Press).
\newblock \doi{10.1017/CBO9780511626265}
\bibAnnoteFile{MR1695775}

\bibitem[{van Lint and Wilson(1992)}]{MR1207813}
van Lint, J.~H. and Wilson, R.~M. (1992).
\newblock \emph{A course in combinatorics} (Cambridge: Cambridge University
  Press)
\bibAnnoteFile{MR1207813}

\bibitem[{Van~Mieghem(2011)}]{MR2767173}
Van~Mieghem, P. (2011).
\newblock \emph{Graph spectra for complex networks} (Cambridge University
  Press, Cambridge)
\bibAnnoteFile{MR2767173}

\bibitem[{Willinger et~al.(2009)Willinger, Alderson, and Doyle}]{MR2509062}
Willinger, W., Alderson, D., and Doyle, J.~C. (2009).
\newblock Mathematics and the {I}nternet: a source of enormous confusion and
  great potential.
\newblock \emph{Notices Amer. Math. Soc.} 56, 586--599
\bibAnnoteFile{MR2509062}

\end{thebibliography}


\end{document}


\onecolumn
\firstpage{1}

\title[Supplementary Material]{{\helveticaitalic{Supplementary Material}}:
\\ \helvetica{Origin of Biomolecular Networks}}

\maketitle

\section{Graph Theory and Its Application to Biomolecular Networks}
In this section, we discuss fundamentals of graphs and networks as well as other important topics that are critical for the study and analysis of biomolecular networks. 

Our discussions deal with an abstraction that facilitates reasoning about a set of entities, denoted $V$ and a binary relation $E \subseteq V \times V$: the binary relation is usually irreflexive,  asymmetric and not necessarily transitive. It is often represented as a directed graph, with vertices $V$ and edges $E$. When $V$ denotes biomolecules and $E$ denotes interactions (e.g., regulations, proximity, binding, synteny, etc.), the resulting graph is called a biomolecular network, object of our study. Such networks evolve over time with additions and deletions to the sets $V$ and $E$.

\section*{Graphs from a Combinatorial Perspective}

We collect here essential results from  graph theory.
For these results, we refer to Serre \cite{serre2} and Biggs \cite{biggs}.

\begin{defn}\label{defn:graph}{\rm A (directed) graph $X$ is a pair of sets $\mathbb{X}=(V, E)$ (the set of vertices $V$ and edges $E$ respectively)}, where a directed edge $e$ is defined as an ordered pair of vertices $e=(a,b)$, where $a$ is the origin of the directed edge $e$ and  $b$ is the terminus of $e$. A rigorous definition then is  augmented with two maps
$$\phi_1:\ \ \ E \rightarrow V\times V:\ \ \ \ \ \ \  \ e
\mapsto (o(e),t(e)),$$ 
and
$$\phi_2:\ E\rightarrow E: \ \ \ \ \ \ \ \ \ \ e\mapsto {\overline e} $$
that  satisfy the following condition: for each $e$,
${\overline {\overline e}}=e$, ${\overline e}\not=e$, and
$t({\overline e})=o(e)$. The bar operation is reversing the direction. Henceforth, we denote the graph $X = (V, E, o, e)$ as a 4-tuple.
\end{defn}

For each edge $e\in E$, $o(e)$ is called the {\it origin} of the edge $e$ and $t(e)$ is called the {\it terminus} of the edge $e$. A graph $Y=(V^\prime,E^\prime, o^\prime, t^\prime)$ is called a subgraph of $X$ if $V^\prime\subseteq V, E^\prime \subseteq E$ and $o^\prime$ and $t^\prime$ are restrictions of $o$ and $t$ respectively to $E^\prime$.

For some studies we need to provide an orientation of an undirected graph. An {\it orientation} of a graph $X$ is a subset $E_{+}$ of the edges such that $E$ is the disjoint union of $E_{+}$ and ${\overline{ E_{+}}}$.

\begin{defn}
{\rm A {\em walk} of length $n$ in a graph is a sequence of alternating vertices and edges, 
\[
\langle v_0, e_1, v_1, e_2, \ldots, v_n, e_n \rangle,
\]
such that $o(e_i)=v_{i-1}$ and $t(e_i)=v_{i}$ for all $i = 1, \ldots, n$.}
\end{defn}


\begin{defn}{\rm A graph $X=(V,E, o, e)$ is said to be connected if any two vertices are the extremities of at least one walk. The maximally connected subgraphs (under the relation of inclusion) are called the connected components of $X$.}
\end{defn}

\begin{defn}\label{defn:bipartite} 
{\rm A graph is called {\it bipartite} if the vertex set can be partitioned into two parts $V_1$ and $V_2$ such that each edge has one
vertex in $V_1$ and one vertex in $V_2$.}
\end{defn}

The distance between two vertices $u$ and $v$ is the length of the shortest walk connecting them, if both vertices are in the same connected component ($\infty$, otherwise). The shortest walk connecting $u$ and $v$ is called a \emph{geodesic}.

Let $n$ be an integer $\geq 1$. Consider the oriented graph on $Z/nZ$,
and the orientation is given by the edges $[i,i+1]$ ($i\in Z/nZ$)
with $o([i,i+1])=i$ and $t([i,i+1])=i+1$.

\begin{defn}\label{defn:circuit}{\rm A subgraph $Y$ of a graph $X$ is called a {\it
 circuit} of length $n$ if it is isomorphic to the circle graph on $Z/n Z$.
 
A circuit of length 1 is called a {\it loop}. If the relation $E$ is irreflexive then the graph is loop-free.
}
\end{defn}
  
\begin{defn}\label{defn:simple}{\rm A graph is called {\it combinatorial or simple} if  it has no circuit of length $\leq 2$.
}
\end{defn}

\begin{defn}\label{defn:tree}
{\rm A non-empty connected graph $T$ without circuits is called a tree.
}
\end{defn}

\begin{defn}\label{defn:weight}
{\rm A weighted  graph $G$ has weights assigned with edges, by the weight function,
$w:V\times V \rightarrow \mathbb{R}$ non-negative, with $w(u,v)=0$ if and only if $e=(u,v)\notin E$. The weighted degree $d_v$ of a vertex $v$ is defined as $d_v:=\sum_u w(v,u)$. We also define volume $V$ of the graph as $vol(G):=\sum_v d_v$.
}
\end{defn}

\section*{Graphs from an Algebraic Perspective }

Certain linear operators can be associated with a graph and can be given a physical meaning in terms of diffusion (of information) over the graph, as common in the signaling games over the biomolecular networks. Spectral analysis of such linear operators yields eigenvalues, eigenvectors, and spectra of graphs, playing important roles in determining various properties of the network -- specifically, with respect to how information diffuses over them (see \cite{MR1421568} and \cite{MR2248695}). 

\begin{defn}
{\rm Algebraically a graph $G$ (network) can be represented as an $n\times n$ adjacency matrix $A(G)$, in which, $A_{ij}$ is 1 iff $\exists e \in E, o(e) = i \;\&\; t(e) = j$; otherwise it is 0. The matrix is symmetric if the graph is undirected, i.e., $e = \bar{e}$, $\forall e \in E$. If the graph $G$ is weighted, then $A_{ij} = w(i, j)$ for every edge $(i,j) = e \in E$, and $0$, otherwise. }
\end{defn}

We can think of $A$ as operating on the space $V=\mathbb{}{C}^n$ of complex $n$-tuples written as column vectors $X$ as follows: $X\rightarrow AX$. $X$ can be thought of as values of a function evaluated on the vertices. One can show that there exist lines through the origin,  in $V$ that are left invariant along those lines. That is to say, there exist scalars $\lambda_i$ (called eigenvalues), and corresponding non-zero vectors $X_i$  (called eigenvectors spanning invariant lines) that span invariant lines  such that $A_i=\lambda_i X$, for $1\leq i\leq n$. The {\em spectrum} of the graph $X$ is defined to be  $Spec(X):=Spec(A):=\{\lambda_1, \cdots, \lambda_n\}$, a collection of $A$'s eigenvalues.

It can also be shown that if $A$ is a real symmetric matrix, then the eigenvalues of $A$ are real and its spectrum can be presented in decreasing order, i.e., $\{\lambda_1 \geq \lambda_2 \geq \cdots \lambda_n \}$. This fact is very important for our study of graphs and networks.

Let us consider a more general weighted graph as defined earlier.
Let $T$ be the diagonal matrix with $d_v$ along the diagonal. First, consider the stochastic matrix
$P = T^{-1} A$, which may be thought of as describing the probabilities of certain ``information'' being moved from one node to a neighboring node by a diffusion process. Let $\{v_0,e_0, v_1, e_1, \cdots, v_s\}$ be a random walk in the graph with $(v_{i-1},v_i) \in E(G)$, for all $1\leq i\leq s$, and determined by transition probabilities $P(u,v)=Prob(x_{i+1}=v|x_i=u)$ which are independent of $i$. Normally we take $p(u,v)=w(u,v)/d_u$, as defined by the stochastic matrix $P$.

Then, let $f:V\rightarrow \mathbb{R}$ with $\sum_v f(v) =1$ be a probability distribution on $V(G)$. Then $\sum_v P(u,v)=1$.  Then for any initial  distribution $f:V\rightarrow \mathbb{R}$ with $\sum_v f(v)=1$, the distribution after $k$ steps is $P^k f$, where $f$ is viewed as a column vector and  $P$ is the matrix of transition probabilities. In particular,  a probability distribution satisfying the fixed point equation $\phi = P \phi = P^2 \phi = \cdots = P^k \phi$ describes the stationary distribution of the diffusion process and can be described as an eigenvector of the corresponding matrix.

Thus intuitively, algebraic techniques allow thinking about the graph features in terms of a set of ``blurrier'' notions such as random walks (instead of walks), diffusion distances (instead of geodesic distances), ranks (instead of informational relevance), etc. However, because such spectral analysis is based on linear algebra, the underlying algorithms become tractable.

The adjacency matrix should be best viewed as an operator on functions of $V(G)$. A modified operator, called the Laplacian operator is the most effective formulation. The Laplacian operator can be used in interpolation on graphs, graph clustering, resistance networks, rapid mixing, linear solving, linear optimization, and many other applications.

Thus, one may define $L = T - A = T (I -P)$, as the {\em Laplacian Matrix} of $G$,  where $L$ is defined as follows: $L(u,v)=-w(u,v)$ (when $u$ and $v$ are distinct), and $d_v$ if $u=v$. Imagine assigning a scalar-valued \emph{rank} function $\rho: V(G) \to \mathbb{R}: v \mapsto \rho(v)$ such that the following \emph{Dirichlet Sum} of $G$
\[
\sum_{(u, v) \in E}
w(u, v) \left( \rho(u) - \rho(v) \right)^2,
\]
is minimized. Thus $\rho$ has the meaning that if a gene in a GRN is important then the genes it regulates and the genes that regulate it are also important; one expects p53 to be labeled as an important gene because of its ``hubbiness,'' but so also, MDM1, ATM, BRCA1, etc. as they are in the pathways directly regulating p53; and also making the genes such as p63 and p73 important as they are regulated by this cluster of genes (which may have preferentially attached themselves to p53 and its duplicates, which they continue to regulate).
Note that solution to the optimization problem for Dirichlet sum  is given by the following equation (under suitable conditions---see \cite{MR3376760, MR2677125, MR1421568, MR2248695}):

\[
\frac{1}{d(x)} \sum_{(y,x) \in E}\left(\rho(x) - w(y, x)\rho(y) \right) = (I - P)\rho = 0.
\]
Thus functions such as $\rho$ can be rapidly computed by iterating over the graph while performing weighted-averaging. An example of this process is seen in Google's PageRank algorithm based on the ``Random Surfer (with Teleportation) Model,''. This and other PageRank algorithms have been successfully  applied 
\emph{mutatis mutandis} to rank genes in a GRN (GeneRank), to rank  Proteins in a PPI Networks (PPIRank), and in other biomolecular networks (see the survey \cite{MR3376760}).

From now, we assume that $G$ is weight symmetric $w(u,v)=w(v,u)$. Then the eigenvalues of $L(G)$ are real, and indeed $0=\lambda_0\leq \lambda_1\leq \cdots \leq \lambda_{n-1}$. Then $\lambda_G:=\lambda_1$ is called the {\em spectral gap} of $G$. The spectral gap (and other eigenvalues) can be determined by the Courant-Fisher theorem. For example, if one considers ${L}$ as an operator on the space of functions $g:V(G)\rightarrow \mathbb{R}$, then
$$\lambda_G:=\lambda_1=\inf_{g\perp {\bf 1}} {\frac{\langle g, {L}\rangle}{\langle g, g \rangle}}$$

 One can show that if the spectral gap $\lambda_G$ is large, and $k$ is large enough any initial distribution $f$ converges to the stationary distribution very rapidly.

\section*{Expansion Properties and Information flows in Graphs}

We shall consider graphs ${X}=(V,E, \cdot, \cdot)$, where $V$ is the set of vertices and $E$ is the set of edges of ${X}$. We will assume that the graph is undirected and connected and we shall only consider finite graphs. For $F\subset V$, the {\it boundary} $\partial F$ is the set of edges connecting $F$ to $V\setminus F$. The {\it expanding constant}, or {\it isoperimetric constant} of ${X}$ is defined as, 
\begin{equation}\label{eqn:Cheeger-constant}
h({X})=\min_{\emptyset\not= F\subset V} {\frac{|\partial F|}{\min\{|F|,|V\setminus F|\}}}
\end{equation}

Moreover if ${X}$ is viewed as the graph of a communication network, then $h({X})$ measures the quality of the network as a transmission network. In all applications, the larger the $h({X})$ the better, so we seek graphs (or families of graphs) with $h({X})$ as large as possible with some fixed parameters.

In \cite{MR752035}, M. Tanner introduced another notation for the expansion coefficient. Let as before ${X}=(V,E, \cdot, \cdot)$, be a graph where $V$ is the set of vertices and $E$ is the set of edges of ${X}$. Let $ X \subseteq V$ with $ |X| \leq {\alpha}|V|$, then
\begin{equation}\label{eqn:Cheeger-constant-alpha}
c({\alpha}) =\min_{\emptyset\not= X\subset V \wedge |X| \leq {\alpha}|V|} \;\; {\frac{|\partial X|} {\min\{|X|,|V\setminus
X|\}}}
\end{equation}

It is well-known that the expansion properties of a graph are closely related to the eigenvalues of the \emph{adjacency matrix} $A$ of the graph ${X}=(V,E)$; it is indexed by pairs of vertices $x,y$ of ${X}$ and $A_{xy}$ is the number of edges between $x$ and $y$. When ${X}$ has $n$ vertices, $A$ has $n$ real eigenvalues, repeated according to multiplicities that we list in decreasing order
\begin{center}
${\lambda}_0 \geq {\lambda}_1\geq \ldots \geq {\lambda}_{n-1}.$
\end{center}
It is also known that if ${X}$ is $D$-regular, i.e. all vertices have degree $D$, then ${\lambda}_0 = D$ and if moreover the graph is connected ${\lambda}_1 < D$. Also ${X}$ is bipartite if and only if $-{\lambda}_0 $ is an eigenvalue of $A$. We recall the following (see for example \cite{MR1421568}   \cite{MR1989434}):
\begin{theorem}
Let ${X}$ be a finite, connected, $D$-regular graph then
$$(D-\lambda_1)/2 \leq h({X})\leq \sqrt{2D(D-\lambda_1)}.$$
\end{theorem}
And
\begin{theorem}(see \cite{MR1421568}, \cite{MR1989434})
Let $({{X}}_m)_{m \geq 1} $ be a family of finite connected, $D$-regular graphs with $|V_m|\to + \infty$ as $m \to \infty$. Then
$$\liminf_{N\rightarrow\infty} \lambda_1({X}_m)\geq 2\sqrt{D-1}.$$
\end{theorem}
This leads to the following.
\begin{defn}
A finite connected, $D$-regular graph ${X}$ is \emph{Ramanujan} if, for every eigenvalue $\lambda$ of $A$ other than $\pm D$, one has $\lambda \leq 2\sqrt{D-1}.$
\end{defn}
We will also need an important definition.

\begin{defn} [Bipartite Ramanujan Graphs]
Let ${X}$ be a $(c,d)$-regular bipartite graph. Then ${X}$ is called a Ramanujan graph if
$$\lambda({X})\leq \sqrt{\left({c}-1\right)}+\sqrt{\left({d}-1\right)}.$$
\end{defn}

It is known that computing the expansion coefficient of arbitrary graphs is an NP-complete problem. Thanks to the work of  Tanner,  and Alon and Millman, one can derive bounds on the expansion coefficient in terms of $\lambda$.  The complexity of determining $\lambda$, though in P, is still  difficult if the number of vertices is large (for example of the order $10^4$-$10^6$ or more for biomolecular networks such as GRN or PPI).

There are useful bounds on $\lambda$ for arbitrary (bipartite) graph $X$ in terms of the number of edges, the maximum degree $\lambda_{max}$, and  the rank $r_\chi$ of the adjacency matrix of $X$. Since effective upper bounds exist on $\lambda_{\max}$, and $r_\chi$, (see \cite {MR2988185}) we thus obtain a bound that is easily computable. 

An expander graph is a  highly connected sparse graph (see, for example \cite{Sarnak:AMS}). Expander graphs have numerous  applications including those in communication science, computer science (especially complexity theory), network design, cryptography, combinatorics  and pure mathematics (see the references under Bibliographic Notes below))). Expander graphs have played a prominent role in recent developments in coding theory (LDPC codes, expander codes, linear time encodable and decodable codes, codes attaining the Zyablov bound with low complexity of decoding (see the Bibliographic notes for references).


\begin{defn}
{\rm A matrix $A$ with rows and columns indexed by a set $X$ is called irreducible when it is not possible to find a proper subset $S$ of $X$ so that $A(x,y)=0$, whenever  $x\in S$ and $y\in X\setminus S$. Equivalently, $A$  is not irreducible if and only if it is possible to apply a simultaneous row and column permutation on $A$ to get a matrix in a square block form so that one of the blocks is a zero block. For the following lemma, see for example (\cite{MR2978290} p. 363).
}
\end{defn}

\begin{lem} Let $ D$ be a finite graph. Then the adjacency matrix of $A$ is irreducible if and only if $D$ is connected.
\end{lem}

We shall also need the following. 

\begin{prop}[Perron-Frobenius]
\label{prop:Perron-Frobenius}
Let $A$ be an irreducible non-negative matrix. Then, there is up to scalar multiples, a unique non-negative eigenvector ${\bf a}:=(a_1,a_2,\cdots,  a_n)$ all of whose coordinates $a_i$ are strictly positive. The corresponding eigenvalue  ${\lambda}_0 $ (called the dominant eigenvalue of $A$) has algebraic multiplicity  1 and ${\lambda}_0 \geq {\lambda}_i$ for any eigenvalue ${\lambda}_i$ of $A$.
\end{prop}

We recall the following special case of Courant-Fisher theorem (also, called the Raleigh-Ritz Theorem) (see for example, (\cite{MR2978290}, Theorem 4.2.2))

\begin{theorem}
Let $A$ be an $n\times n$ Hermitian matrix over the complex field  ${C}$, then it is known that all its eigenvalues are real, with maximum eigenvalue ${\lambda}_{\max}$ (i.e. the spectral radius of $A$ ). For ${\bf 0}\not=X\in {C}^n$,  define the Raleigh quotient
$$R_X:= {\frac{{X^*}^T A X}
{ {{X^*}^T X} }}.$$
Then  ${\lambda}_{\max}=\max_{X\not={\bf 0}} R_X$. Furthermore, $R_X\leq {\lambda}_{\max}$ with equality if and only if $X$ is an eigenvector corresponding to the eigenvalue  ${\lambda}_{\max}$.
\end{theorem}




\section*{Groups Acting on Graphs}

To understand symmetries of graphs, and to understand original motivation of graphs as objects associated with topology and algebraic topology, we briefly discuss groups acting on graphs (as groups acting on topological  spaces). This converts graph isomorphism problems into much more manageable group isomorphism problem. It is needed in alignment and motif detection, We say that a group $G$ {\it acts} on a graph $X(V,E,o,t)$ if it acts on $V$ and $E$ such that the actions are compatible with $\phi_1$ and $\phi_2$, i.e., it commutes with $\phi_1$ and $\phi_2$, i.e., $g(\phi_1(e))=\phi_1(g(e))$, $o(g(v))=g(o(v))$,
$t(g(v))=g(t(v))$ and $g(\phi_2(e))=\phi_2(g(e))$, $\forall g \in G$.

An {\it inversion} is a pair consisting of an element $g\in G$ and an edge $e\in E$ such that $g(e)={\overline e}$. We will say that a group acts {\it freely } on $X$ if it acts without inversion and $g=1$ is the only element having a fixed point. For the following result, we refer to Serre \cite[Page 27]{serre2}.

\begin{theorem}\label{thm:tree} If $G$ acts freely on a tree, then $G$  is a free group.
\end{theorem}

\begin{defn}\label{defn:cayley}{\rm {\em The Graph $X(G,S)$}: {
\rm Let $G$ be a group and $S\subseteq G$. The {\it Cayley graph} $X(G,S)$ is defined as the oriented  graph with the vertex set  $G$ and  edge set $E=G\times S$.
}}
$$ o(g,s)=g\ \ \ {\rm and}\ \ \ t(g,s)=gs $$ 
\end{defn}

The group $G$ acts on $X(G,S)$ by left multiplication. This action preserves 
orientation; and  its action is free on the set of vertices and on the set of edges.

\begin{prop}\label{prop:graph:prop}
\begin{enumerate}
\item[i)] $X(G,S)$ is connected if and only if $S$ generates $G$. In fact, the connected components correspond in a 1-1 fashion to the cosets of $H=\langle S\rangle$ (the group generated by $S$).
\item[ii)] $X$ contains a loop  if and only if $1\in S$.
\item[iii)] For $X$ to be combinatorial, it is necessary and sufficient that \hfill\break
$S\cap S^{-1}=\emptyset$.
\end{enumerate}
\end{prop}



The adjacency matrix provides significant amount of information about the graph. For example $A^r$ gives the number of walks of length $r$ between vertices.


\begin{defn}\label{defn:girth} 
{\rm The {\it girth} of $X$ is the smallest positive integer $r$ such that $Trace(A^r)>0$. Let $d(X)$ be the smallest integer (if it exists) so that for every pair of vertices $(u,v)$ there is a walk of length at most $d$ from $u$ to $v$. Then $d(X)$ is called the {\it diameter} of the graph $X$.
}
\end{defn}

Important results about diameter, girth and other combinatorial invariants (important for biomolecular networks) and bounds on them in terms of spectral invariants can be found for 
general graphs in  Bollobas (\cite{bollobas1}, pp. 156--157).

\begin{lem}\label{lem:eigenvalues} Let $X$ be a $k$-regular graph. Then
\begin{enumerate}
\item[i)]  $\lambda_i\leq k$, for $0\leq i\leq n-1$.
\item[ii)] $\lambda_0=k$ and $m(\lambda_0)$ equals the number of connected components of $X$.
\item[iii)] $\lambda_{n-1}=-k$ if and only if $X$ is bipartite.
\item[iv)] For a bipartite graph $X$, if $\lambda$ is an eigenvalue
with multiplicity $m (\lambda)$, then so is $-\lambda$ with multiplicity $m(\lambda)$.
\end{enumerate}
\end{lem}




The adjacency matrix $A$ can be considered as the matrix of a Hecke operator on $l^2(X)$ (which can be called the adjacency operator) as $A(f(x))=\sum_{y\in V}  A(x,y) f(y)$ (here $A(x,y)=1$ if there is an edge from $x$ to $y$ and 0 otherwise). As mentioned earlier in the context of rank function, another interpretation of the adjacency operator is an averaging function of information  contained on the neighboring vertices that flow along the adjacencies. An iterative process, then leads to mixing and diffusion globally in the network via the adjacency or the corresponding Laplacian operators.




\section*{Miscellaneous Related Comments}

\paragraph*{Remark (1)}
The topics we expanded upon are the following (in order): (1) graphical representation of Networks; (2) algebraic representation of graphs; and finally, (3) algebraic properties, such as spectral analysis that provide us large number of tools and techniques to deduce various properties of the biomolecular networks. For complex and mid-size networks and models, there are important algorithmic questions related to random graphs and their evolution (Erd\"os-Renyi model), degree-distribution, power laws, preferential attachment models, scale free networks, random-walks and mixing, spectral distance, graph similarity, clustering, smoothing analysis, sparsification and linear solvers and applications.







\paragraph*{Remark (2)}
Algebraic representation of biomolecular networks has several advantages. In particular, the combinatorial meaning is inherent in the Adjacency matrix $A(X)$  of a graph $X$.  In the study of biomolecular networks, it is very important to know the number of walks  of length $m$ between any two entities $i$ to $j$,  $w_{ij}^m$,  (e.g. between any two proteins in the  PPI network). But that is given by the $(ij)^{\rm th}$ entry in $A^m$.  From this one can deduce that the diameter of the network is the dimension of what is called the adjacency algebra $A(L)$ associated with $A$. If this diameter is small, we can deduce that the networks shows small world phenomenon, as we explained in the main text. The sequence ($w_{ij}^m)$ can be put into coefficients of a series called the zeta function $\zeta(X)$ of the graph $X$, and it has a very simple form as a rational polynomial in terms of what are called spectra (or frequencies) of the network. And from $\zeta(X)$ we can find simple expressions for the set of walks.

\paragraph*{Remark (3)} One can explain spectra (frequencies) of a biomolecular network in an intuitive manner as follows. In analogy with the physical sciences, if one considers the whole network as a space, one can think of its vibration modes, that is the frequencies (eigenvalue spectra) and their amplitudes (corresponding eigenvectors). One can indeed determine the shape of the the biomolecular network from its Spectra (in the manner of Marc Kac (Can one determine the shape of the drums?  (from its spectra)~\cite{MR0201237} ), and its analogy to networks and graphs \cite{Chung-Shape}. The advanced matrix of the biomolecular network can be then shown to take a simple diagonal form with respect to the basis of eigenvectors, and that leads to immense simplification of explanation of many phenomena associated with the biomolecular networks, and plays important roles in the spectral algorithms and their complexity analysis.  The coefficients of  polynomials whose roots are the eigenvalues, carries information about the motifs at the nodes.  The spectral gap $\lambda(X)$  (the difference  between the two largest frequencies) of the biomolecular network determines expansion in the biomolecular network, meaning how fast spreading and  mixing takes places between a set of nodes. How fast partitioning and cuts can be carried out----important ingredient of many algorithms in biomolecular networks. In particular the spectral gap of the biomolecular network, $\lambda(X)$,  gives lower bounds in terms of the combinatorial invariant, the Cheeger constant, as defined below. We will state isoperimetric inequality results where $\lambda(X)$ provides an excellent lower bound on $h(X)$, and for explanation of combinatorial phenomenon or for spectral approximation to combinatorial algorithms, $\lambda(X)$ can replace $h(X)$ and still one can get excellent approximate results. For example, it is very easy to show fast mixing and spreading phenomenon----it is just $A^n f=\lambda(X)^n f$, where $f$ is the initial distribution on the nodes.

\paragraph*{Remark (4)}
We have defined important algebraic concepts in graph theory in precise mathematical terms. Networks (molecular or otherwise) have algebraic representations as their adjacency matrix (from which one can re-derive the graphical representation if one wishes). From these one derives algebraic invariants such as eigenvalues and spectra of graphs, eigenvectors, and spectral gaps. In the main text we had defined the combinatorial invariant called the isoperimetric constant $h(G)$. The determination of $h(G)$ is an NP-complete problem. In the next sections we will show that $h(G)$ can be bounded in terms of the spectral gap $\lambda(X)$. This helps immensely in the determination of spectral graph algorithms as approximation to NP-complete combinatorial algorithms.





\paragraph*{Remark (5)} 
One application of large spectral gap is rapid mixing (one can envisage that it is expected to  prove very important in many biological applications). Indeed, the  discrete analog of Cheeger inequality has increasingly crucial utility  in the study of random walks and rapid mixing on Markov chains  and new powerful spectral  techniques such as Heat-Kernels and Sobolov inequalities have emerged to deal with general graphs (see \cite{MR2248695}, \cite{MR2108973}). Rapid mixing in Markov chains can be framed as: How long does it take for an irreducible finite state Markov chain to converge to equilibrium?
A fundamental application is to Markov Chain Monte Carlo  (MCMC) simulation algorithms that  are used widely in the scientific community to simulate Gibb’s measures and to derive approximate solutions to difficult combinatorial questions. as Markov Chain Monte Carlo simulations are used widely in the scientific community to simulate Gibb’s measures and to derive approximate solutions to difficult combinatorial questions----that have  high complexity and many of which appear in the topological analysis of biomolecular networks (such as clustering, community detection, and so on). Indeed two of the most heavily studied problems in the analysis of networks are graph  clustering and graph diffusion.
We already studied the importance graph clustering biomolecular networks in the main text. Graph diffusion refers to problems involving spreading or propagation along the edges of a graph.
These problems are of fundamental important in algorithms such as PageRank and Hitts algorithms (see \cite{MR2677125}, \cite{MR3726904},\cite{MR2248695}, and \cite{MR2108973}).

\section{Bibliographic Notes}


For further comprehensive treatment of applications of {\bf spectral graph theory} to biomolecular networks we refer to \cite{MR2108973} and the survey article \cite{MR2527959}. For state of the art in spectral methods in algorithmic analysis, we refer to the lecture notes of Spielman (http://www.cs.yale.edu/homes/spielman)  biomolecular networks algorithms in  clustering, mixing, partitioning, random walks, Schur complements, effective resistance and applications, expander graphs and applications, graph sparsification and related algorithms, testing isomorphism of networks.

Related to spectral graph theory, {\bf expander graphs} have become prominent in many recent developments in information and coding theory (LDPC codes, expander codes, linear time encodable and decodable codes, codes attaining the Zyablov bound with low complexity of decoding (see \cite{MR752035}, \cite{MR1465731}, \cite{MR1989434}, \cite{MR1102679}, \cite{MR1308046}, \cite{MR2988185}, \cite{MR2580853}, \cite{MR2042419}, \cite{MR1959169}, \cite{MR1676451}, \cite{MR2869010}),\cite{MR1465731} \cite{MR1465732}, \cite{Tanner:IEEE:2001}, \cite{MR1959169},  \cite{Barg:Zemor}, \cite{Guruswamy:Indyk}, and others). 

The following articles are relevant for a comprehensive treatment of {\bf spectral graph theory}, and {\bf spectral graph  theoretic algorithms} that are relevant to biomolecular networks described in the main text: \cite {MR1140703}, \cite {MR546860}, \cite {MR972207}, \cite {MR506522}, \cite {MR840466}, \cite{MR1141921}, \cite {MR1002568}, \cite{MR1141922},  \cite{MR928566}, \cite {MR572262}, \cite {MR926481},  \cite {MR964069}, \cite{MR1141927},\cite {MR37},\cite {MR3600878},
 \cite {MR3147592},\cite {MR2988185}, \cite {MR2580853},\cite {MR1959169},\cite {MR1403868},\cite {MR1359909}, \cite {MR1079013},  \cite {MR1390759}, \cite {MR1207813}, \cite {MR1308046}, \cite {MR963118}, \cite {MR0465510}, \cite {MR939574},  \cite {MR451347}, \cite {MR1127181},\cite {MR1259014},\cite {MR1081951}, \cite {MR1102679}, \cite {MR1465731}, \cite {MR752035}, \cite {MR1695775}).

For {\bf application of spectral graph theory} to networks (e.g., biomolecular) we suggest \cite{MR2767173}, \cite {MR3730470}, \cite {MR1673501}, \cite {MR2759498}, \cite {MR2759501}, \cite {MR2108978} \cite {MR2982456}, \cite {MR2527959}, \cite {MR2468898}, \cite {MR2509062}, \cite {MR2857452}, \cite {MR2348125},  \cite {MR2780129}), and general {\bf spectral analysis} applicable to biomolecular networks (\cite{MR2798107}, \cite{MR2560260}, \cite{MR2304986}, \cite{MR2248695}, \cite{MR2108973}.

For {\bf algorithmic  analysis} of various properties of biomolecular networks discussed below we refer to  (\cite{MR2767173}, \cite {MR3730470}, \cite {MR1673501}, \cite {MR2759498}, \cite {MR2759501}, \cite {MR2108978} \cite {MR2982456}, \cite {MR2527959}, \cite {MR2468898}, \cite {MR2509062}, \cite {MR2857452}, \cite {MR2348125},  \cite {MR2780129}), and general spectral analysis applicable to biomolecular networks (\cite{MR2798107}, \cite{MR2560260}, \cite{MR2304986}, \cite{MR2248695}, \cite{MR2108973}).
\newpage




\bibliographystyle{frontiersinSCNS_ENG_HUMS}
\bibliography{bibliography}